\documentclass[10pt, aps,prc,twocolumn,superscriptaddress,preprintnumbers,
amsmath, 
floatfix,
longbibliography,
nofootinbib
]{revtex4-1}
\usepackage[T1]{fontenc}
\usepackage[utf8x]{inputenc} 
\usepackage{adjustbox}          
\usepackage[caption=false]{subfig}
\usepackage{url}
\usepackage{color}
\usepackage{float}
\usepackage[pdftex,colorlinks=true, linkcolor = blue, citecolor=blue,urlcolor=blue, bookmarksnumbered=true, bookmarksopen=true]{hyperref}
\usepackage{longtable}
\usepackage{amsfonts}
\usepackage{wrapfig,bm,bbm}

\usepackage[normalem]{ulem}

\newcommand{\beq}{\begin{equation}}
\newcommand{\eeq}{\end{equation}}
\newcommand{\bea}{\begin{eqnarray}}
\newcommand{\eea}{\end{eqnarray}}
\newcommand{\Tr}{\mathrm{Tr}}

\newcommand{\eF}{\varepsilon_{F}}

\begin{document}

\title{Sensitivity to the initial conditions of the Time-Dependent Density Functional Theory} 

\author{Aurel Bulgac}
\affiliation{Department of Physics, University of Washington, Seattle, WA 98195--1560, USA}
\author{Ibrahim Abdurrahman}
\affiliation{Department of Physics, University of Washington, Seattle, WA 98195--1560, USA}  
\author{Gabriel Wlaz\l{}owski}
\affiliation{Faculty of Physics, Warsaw University of Technology, 
  Ulica Koszykowa 75, 00--662 Warsaw, POLAND}
\affiliation{Department of Physics, University of Washington, Seattle, WA 98195--1560, USA}

\date{\today}

\begin{abstract}

Time-Dependent Density Functional Theory is mathematically
formulated through non-linear coupled time-dependent 3-dimensional
partial differential equations and it is natural to expect a strong
sensitivity of its solutions to variations of the initial conditions,
akin to the butterfly effect ubiquitous in classical dynamics. Since
the Schr\"odinger equation for an interacting many-body system is
however linear and mathematically the exact equations of the Density
Functional Theory reproduce the corresponding one-body properties, it
would follow that the Lyapunov exponents are also vanishing within a
Density Functional Theory framework.  Whether for realistic
implementations of the Time-Dependent Density Functional Theory the
question of absence of the butterfly effect and whether the dynamics
provided is indeed a predictable theory was never discussed.  At the
same time, since the time-dependent density functional theory is
a unique tool allowing us the study of non-equilibrium
dynamics of strongly interacting many-fermion systems, the question of
predictability of this theoretical framework is of paramount
importance.  Our analysis, for a number of quantum superfluid many-body systems
(unitary Fermi gas, nuclear fission, and heavy-ion collisions)
with a classical equivalent number of degrees of freedom ${\cal O}(10^{10})$ 
and larger, suggests that its maximum Lyapunov exponents are negligible for all 
practical purposes.

\end{abstract}

\preprint{NT@UW-21-06}

\maketitle

\section{Preamble}\label{sec:pre}

Dynamical systems are often deterministic, but at the same time
unpredictable in the long run.  In 1884 G\"osta Mittag-Leffler
initiated the idea of a new international prize in mathematics, in
honor of King Oscar II of Sweden and Norway. The first challenge was
to solve the $N$-body problem interacting according to Newton's law of
gravitation.  In 1888 Henri Poicanr\'e submitted his answer, about 300
of hand-written pages, to which upon request he added another 100 or so
hand-written pages, and he was awarded the prize. The initial
Poincar\'e's ``solution'' had an error, he missed what we now know as
the butterfly effect, an error which he fixed before results were
later published~\cite{Poincare:1890}. The entire dramatic story is
available online~\cite{mittag}.  Poincar\'e's insights were further
extended by the work of Lyapunov~\cite{Lyapunov:1992}, who introduced
the now ubiquitous Lyapunov exponents. The topic remained practically
dormant until the paper of \textcite{Lorenz:1963}, who soon after
apparently introduced the meme butterfly effect as a poetic
illustration of the high sensitivity of deterministic systems to
initial conditions.  This is how we have learned that the equations of
motion of a physical system can be deterministic and at the same time
can be unpredictable in the long run. The chaos theory was born,
before anyone had any clue that quantum mechanics would quite soon
change the world in so many ways and in particular our interpretation
of chaos.  Recently a statistical approach was suggested as an
alternative reformulation of the three-body problem
challenge~\cite{Kol:2021,Manwadkar:2020,Manwadkar:2021,Ginat:2021},
which renders this problem deterministic.

Classical chaos studies show that a system of interacting particles, being 
nonlinear in character, would typically show sensitivity to initial conditions. 
Unless a classical Hamiltonian system is integrable it is
characterized by the presence of at least one positive Lyapunov
exponent, rendering any long-time predictions totally unreliable, even
though the equations of motion are unquestionably
deterministic. Popularly this is referred to as the butterfly effect
and in scientific literature as deterministic chaos.
If the time-dependent Density Functional Theory (TDDFT), which by 
default is a nonlinear theory, unlike the many-body Schr{\" o}dinger equation, 
would be sensitive to initial conditions, the TDDFT would be unpredictable  
and useless as a theoretical tool. The sensitivity of a many-body system 
to initial conditions would be equivalent to the absence of the linear regime, 
as any small initial perturbation would increase exponentially in time and the 
many-body system would be basically unstable. 

Our goal is to ascertain whether the TDDFT equations  
are characterized by non-vanishing Lyapunov exponents.   

\section{Quantum Mechanics versus Classical Mechanics} \label{sec:qmcm}

In this section we will review a number of aspects of chaos theory in the
classical and quantum descriptions of nature, which can be skipped by
the informed reader.\\

The Lyapunov exponent $\lambda$ of a dynamical system is a quantity
which characterizes the rate of separation of initially
infinitesimally close trajectories
\begin{align}
\lambda = \lim_{t\rightarrow \infty} \lambda(t) =
 \lim_{t\rightarrow \infty} \frac{1}{t} \ln \frac{|\delta {\bm Z}(t)|}{|\delta{\bm Z}(0)|}
 \label{eq:Z}
\end{align}
where $\delta {\bm Z}(t)$ are the differences between the two full
sets of phase space variables.  In practice one evaluates $\lambda$
via Eq.~\eqref{eq:Z} by starting from a fixed small $|\delta {\bm
Z}(0)|$ and by evolving in time the system for a relatively long time.
For an $N$-dimensional ${\bm Z}(t)$ one defines a spectrum of $N$
Lyapunov exponents.  In the case of a Hamiltonian dynamics, Liouville's
theorem implies that
\begin{align}
\sum_{k=1}^N \lambda_k =0
\end{align}
 and moreover that a number of these Lyapunov exponents identically
vanish in the presence of symmetries and conserved quantities.  In the
case of dissipative dynamics $\sum_{k=1}^N \lambda_k <0$. The sum of
all positive Lyapunov exponents gives an estimate of the
Kolmogorov-Sinai entropy~\cite{Pesin:1977}. A related quantity is the
Lyapunov dimension or Kaplan-Yorke dimension~\cite{Kaplan:1979}.
 
 A particular
class of classical systems is interesting in connection with quantum
dynamics: for any quadratic Hamiltonian in the canonical coordinates
and momenta all the Lyapunov exponents are identically zero. It is
trivial to show that the non-relativistic quantum dynamics with
instantaneous interactions between particles is mathematically
equivalent to a classical system of infinitely many coupled harmonic
oscillators. If a quantum $N$-body system with coordinates $x=({\bm
r}_1,\dots,{\bm r}_N)$ satisfies the time-dependent Schr{\" o}dinger
equation
\begin{align}
i\hbar \dot{\Psi}(x,t)= \int \!\! dy \, H(x,y)\Psi(y,t),
\end{align}
then one can introduce the continuous classical canonical coordinates
and momenta, similarly to fluid dynamics~\cite{LL6:1966}, and the
corresponding total Hamiltonian of the ``classical'' continuum
canonical coordinates.  We suppressed, for the sake of simplicity,
the spin, isospin and any other discrete degree-of-freedom (DoF).  A
set of continuous canonical coordinates and momenta convenient 
for our purpose is
\begin{align} 
&q_x(t) ={\sqrt{2}}\, \text{Re}\,\Psi(x,t), \\ 
&p_x(t)= {\sqrt{2}} \,\text{Im}\,\Psi(x,t), \\
&{\cal H} = \frac{1}{\hbar}\int \!\! dx dy\, \Psi^*(x,t)H(x,y)\Psi(y,t)
\end{align}
where ${\cal H}$ is the corresponding classical Hamiltonian of quadratic form,
and where $x$ labels these canonical coordinates. Using the classical 
Poisson brackets one immediately obtains the classical Hamiltonian 
form of the Schr{\"o}dinger equation
\begin{align}
&\dot{q}_x(t) = \{q_x(t),{\cal H}\}= \;\;\;\frac{\delta {\cal H}}{\delta p_x(t)},\label{eq:Hq}\\
&\dot{p}_x(t) = \{p_x(t),{\cal H}\}= - \frac{\delta {\cal H}}{\delta q_x(t)}.\label{eq:Hp}
\end{align} 
This Hamiltonian form of the Schr{\"o}dinger equation naturally
implies that all Lyapunov exponents of any non-relativistic many-body
Hamiltonian with instantaneous interactions identically vanish, and
thus it strictly follows that there is no quantum chaos, or there is
no sensitivity to the initial conditions in for any quantum many-body
problem with instantaneous interactions. Alternatively, one can state that the Schr{\" o}dinger equation is linear and thus  
no exponential divergence between initially close quantum states can be observed.

This is the reason why for many decades now theorists tried to answer
such questions, using argumentation based on the correspondence
principle: Why a quantum Hamiltonian, which in the classical limit
$\hbar \rightarrow 0$ has a chaotic dynamics, does not shows any
sensitivity to initial conditions? How can one identify the signatures
of chaotic features present in the classical limit in the quantum
description? Most of the studies of quantum chaos focused on several
aspects: i) the presence of either the Poisson or of the random matrix
energy level distributions in quantum systems~\cite{Mehta:1991} 
compared with their corresponding integrable or chaotic classical
counterparts~\cite{Bohigas:1984}; ii) the ``scars'' on the quantum
wave functions in the limit $\hbar \rightarrow 0$ left by the
classical periodic
orbits~\cite{Balian:1974,Berry:1976,Gutzwiller:1977,Berry:1981,Heller:1984}. 
Unfortunately the wave functions ``scars'' cannot be studied in case of 
quantum many-body systems, for obvious reasons. The
role of the boundary effects versus the role of the interaction
between particles were not always clearly separated in such studies.
In an overwhelming fraction of studies of quantum chaos the object was
devoted to billiards, typically in 2 dimensions.  In the present study
we concentrate solely on the role of the interaction between
particles.

In spite of all the insight gathered during the past several decades
in quantum chaos, one did not really put in evidence a sensitivity to
initial conditions in an as clear manner as in classical
dynamical systems.  Recently a new approach has been advocated, the
study of the so called out-of-time ordered correlator
(OTOC)~\cite{Larkin:1969,Maldacena:2016}, in particular the quantity
\begin{align}
C(t)=-\frac{\text{Tr}\left \{e^{-\beta \hat{H}} [\hat{x}(t),\hat{p}(0)]^2\right \}}{\hbar^2\text{Tr}\left \{e^{-\beta \hat{H}}\right \}}, \label{eq:C}
\end{align}
where $[\hat{x}(t),\hat{p}(0)]$ is the commutator of the Heisenberg
coordinate and momentum operators
\begin{align}
& \hat{x}(t)=e^{-i\hat{H}t/\hbar }\hat{x}(0)e^{i\hat{H}t/\hbar },\\  
& \hat{p}(t)=e^{-i\hat{H}t/\hbar }\hat{p}(0)e^{i\hat{H}t/\hbar },\\
&[ \hat{x}(t),\hat{p}(t) ]=i\hbar, \quad [ \hat{x}(t), \hat{p}(0) ]=i\hbar \frac{ \partial \hat{x}(t) }{ \partial \hat{x}(0) },
\end{align}
where $t$ is time and $\beta=1/T$ in the inverse temperature. (It is
not a requirement to use the canonical average.)  The reason for such
a choice is that the Poisson bracket $\{A,B\}$ of the classical limit
of the quantum operators $\hat{A}, \hat{B}$ is related to the quantum
commutator
\begin{align}\label{eq:Lya}
\lim_{\hbar\rightarrow 0} \frac{ [ \hat{A}, \hat{B }] }{ i\hbar } = \{A,B\}.
\end{align}
One can establish an apparent link between the OTOC and classical
chaos, and show the maximum Lyapunov exponent of a quantum system has
an upper bound.  \textcite{Maldacena:2016,Tsuji:2018} have proven
that under rather some very reasonable assumptions
\begin{align}
\lambda_\text{MSS} =\lim_{t\rightarrow \infty} \frac{\ln C(t)}{2t} \leq \frac{\pi T}{\hbar}. \label{eq:lam}
\end{align}
In Ref.~\cite{Maldacena:2016} the authors overlooked that $\{ x(t),
p(0) \}\equiv \partial x(t)/\partial x(0) \propto \exp(\lambda t)$ and thus
$C(t)\propto \exp(2\lambda t)$.  In the classical limit one discusses
the behavior of $\Delta x_k(t)/\Delta x_k(0) \propto \exp(\lambda t)$,
but not of its square $|\Delta x_k(t)/\Delta x_k(0)|^2\propto
\exp(2\lambda t)$ as it is done using the OTOC for quantum mechanical
systems, see Eq.~\eqref{eq:C}.

It was shown that the quantum Lyapunov 
exponent, see Eq.~\eqref{eq:C}, can be  extracted 
from the thermally averaged Loschmidt echo signal~\cite{Yan:2020a},
\begin{align}
L(t) = |\langle \psi | e^{iH_0t/\hbar}e^{-i(H_0+V)t/\hbar}|\psi\rangle |^2, \label{eq:L}
\end{align}
which exhibits a
fast exponential decay $\propto\exp(-2\lambda t)$. 
Here $V$ stands for a small perturbation -- a diagnostic widely used
theoretically and experimentally.
There is thus a
difference between the sensitivity discussed in classical chaos and
quantum chaos. In classical systems one perturbs the initial
conditions, while in quantum studies one often perturbs the
evolution Hamiltonian. Recently, it was argued again, using 
quantum computing reasoning, that there is no butterfly effect 
in a genuine quantum evolution~\cite{Yan:2020}, as was 
argued also for decades in literature.

These various conclusions deserve some introspection.
\begin{itemize}

\item Classically a system manifests a chaotic behavior in isolation,
not in contact with a thermal bath, as \textcite{Maldacena:2016} have
assumed. One can however relate the temperature $T$ used by
\textcite{Maldacena:2016} with the energy level density of a (large)
quantum system
\begin{align}
\rho(E) =\text{Tr}\, \delta (E-H)\propto e^{S(E)},
\end{align}  
where $S(E)$ is the entropy and then the temperature can be defined as
\begin{align}
\frac{1}{T} = \frac{dS(E)}{dE}
\end{align}
for a large isolated system, which reached statistical
equilibrium~\cite{Landau:1980}.

The statistical mechanical equilibrium is not an instantaneous
quality, it is difficult to determine from an instantaneous snapshot of the
system~\cite{Landau:1980}.  The equilibrium characterizes the behavior
of a system, in particular of an isolated system, over a very long
time, when according to Boltzmann the time average becomes identical
to the phase space average.  Thus the time should be long enough for
the system to have managed to cover a relevant part of the phase
space, for the system to be capable of revealing its statistical
equilibrium properties.  The equilibration time for a many-body system
(assuming extensivity of the system properties) is in theory
exponentially large $\propto e^{ \# N}$ as only for a very long time
the many-body density matrix of the system reaches equilibrium as a
result of visiting all allowed phase space cells.  On the other hand,
simple quantities, such as the one-body distribution or the two-body
correlations reach equilibrium in a much shorter time, typically after
a relatively small number of interparticle collisions. An
interparticle collision can be characterized qualitatively only in a
relatively dilute system, and therefore the analysis of dense systems
is a qualitatively complicated issue. In dense systems one can replace
the particles with quasiparticles and perform a similar analysis.  In
particular, the entropy increase in time, often discussed in
literature, should be evaluated over time intervals longer than the
equilibration times.

Eq.~\eqref{eq:C} implicitly assumes that $C(t)$ is calculated for a
system in thermal equilibrium at all times.  Since $C(t)$ increases
exponentially, the inverse rate
\begin{align}
1/\lambda_\text{MSS} \gg \tau_\text{equil}\label{equ:equil}
\end{align} 
should be longer than the equilibration rate of the system.

\item The classical mechanics appears as a singular limit of the
quantum mechanics, and very likely the order in which the limits
$\hbar\rightarrow 0$ and $t\rightarrow \infty$ are taken influence the
interpretation of the results of various studies.

In the classical limit there is no upper limit on the maximum value of
a Lyapunov exponent, which in a sense explains the abundance of
classical chaotic systems, if one is to take the limit
$\hbar\rightarrow 0$ first in Eq.~\eqref{eq:lam}.  At the same time one can justify why
many-body quantum systems do not have random matrix fluctuations on
energy scales $\Delta E$ much larger than the average level separation
in large system~\cite{Brink:1979,Bulgac:1998,Kusnezov:1999}
\begin{align}
\Delta E\propto T^{3/2}\gg \frac{1}{\rho(E_0)}\approx \exp\left (-\frac{aT}{2}\right ), \label{eq:relax}
\end{align}
when statistical equilibrium is achieved. This estimate was obtained by 
considering the behavior of the level density of nonrelativistic many 
fermion systems derived by \textcite{Bethe:1936}
\begin{align}
\rho(E) \propto \exp (\sqrt{a E})
\end{align}
and the corresponding average equilibrium energy and variance
\begin{align}
E_0 \approx aT^2 \quad  \Delta E \propto T^{3/2},
\end{align}
and $a\approx A/10$, where $A$ is the nucleus atomic mass.

\item A quantum system can be initialized with any energy $E$ with an
energy width $\Delta E \ll E$.  Such a packet can likely be chosen to
be also rather well localized so as $\langle \Delta x_k(0)\rangle
\langle \Delta p_k(0)\rangle \sim \hbar$, where $k=1,\ldots, N$.  It
is sufficient to assume that localization is in phase space, and, as
in the case of squeezed states, $\langle \Delta x_k(0)\rangle \propto
\hbar^\alpha$ and $\langle \Delta p_k(0) \rangle \propto
\hbar^{1-\alpha}$ or consider any other orientation of such an ellipse
or even any other shape with comparable area in the phase space.  The
more recent hypothesis of eigenstate thermalization and related
studies~\cite{Deutsch:1991,Srednicki:1994,Zelevinsky:1996} are similar
approaches to define the relaxation time to reach statistical
equilibrium in accordance to the principles of statistical
mechanics~\cite{Landau:1980}.

\item Quantum packets spread with time and if the system is finite in
space, an initial wave packet with a spatial size smaller than the
system size eventually spreads over the entire system, during the
Ehrenfest time $\tau_E$.  For a chaotic system the Ehrenfest time is
determined by $\langle \Delta x_k(0)\rangle \exp(\lambda \tau_E) \sim
L$, if one considers the separation of various trajectories in the
real space. Consider all possible initial conditions of a classical 
Hamiltonian system in a small volume in phase space of compact 
size (linear dimension $L$ in all directions). If  a Hamiltonian system is 
chaotic and various trajectories diverge exponentially, and since 
according to Liouville theorem the phase space volume is conserved, 
with time this initial compact phase space volume evolves into a 
fractal, which can overlap significant parts of the entire phase space.

\item One can prepare two identical systems with slightly different
initial conditions ${\bm Z}_{1,2}(0)$ and record as a function of time
$\Delta {\bm Z}(t)={\bm Z}_1(t)-{\bm Z}_2(t)$ and thus extract their
relative velocity.

Obviously, the absence of the speed of light in Eq.~\eqref{eq:lam}
restricts the validity of this upper bound to non-relativistic systems
only.  Since two initially infinitesimally close trajectories cannot
separate with a relative speed greater than twice the speed of light,
the Lyapunov exponents are strictly vanishing in relativistic
theories. Since the linear momenta however can increase indefinitely,
the character (and maybe even existence) of chaos in a relativistic
system might be qualitatively different from that in a Newtonian
system.  However, if one would still find a butterfly effect in a
relativistic system, when one or a small number of momenta can
increase exponentially, a fact that will point likely to the absence of
ergodicity. A significant amount of energy will then get concentrated 
among few degrees of freedom, which will correspond to a relatively 
small part of the phase space.

\end{itemize}

\section{Strongly interacting quantum many-body systems}\label{sec:qmbs}

The main issue we will be concerned within this work is the presence
or absence of chaotic behavior in a quantum many-body system, under
realistic conditions.
\begin{itemize}

\item When analyzing a concrete quantum many-body system and well
specified phenomena we will assume that the role of $\hbar\neq 0$
cannot be ignored and therefore we will not consider the $\hbar
\rightarrow 0$ limit.

\item For specific physical phenomena often there is no meaning to
consider the limit $t\rightarrow \infty$.  A typical case is a
chemical reaction $AB + C \rightarrow A + BC$ at the later times,
after the complex $BC$ has been formed and it is safely separated from
$A$. However, by slightly changing the initial conditions the emerging 
molecules might end up in drastically different configurations, and this 
can be interpreted as a significant sensitivity to initial conditions..

\end{itemize}

It has been established that for interacting quantum many-body systems
there exist two equivalent quantum mechanical formulations: the
many-body Schr{\" o}dinger equation and the Density Functional Theory
(DFT)~\cite{Hohenberg:1964,Kohn:1965fk,Runge:1984,Kohn:1999fk,Dreizler:1990lr,Gross:2006,Gross:2012},
if one is interested only in the behavior of the one-body or the
number density of the system $n({\bf r},t)$ and the total energy of
the system. However the two formulations, while mathematically proven
to be equivalent, have completely different realizations in terms of
underlying equations. While the Schr{\" o}dinger equation is linear
and is always equivalent to an infinite continuum system of coupled
classical oscillators, the DFT is manifestly a non-linear theory.
Since typical implementations of DFT are formulated in terms of
single-particle wave functions (spwfs), these spwfs can be in the same
manner split into their real and imaginary parts and DFT can be
shown to be mathematically equivalent to very complex continuum
classical system, governed by non-linear (non-dissipative) classical
Hamiltonian equations of motion, similar to
Eqs.~(\ref{eq:Hq},\ref{eq:Hp}).

The equivalence of the many-body Schr{\" o}dinger equation to the 
TDDFT was proven under a number of assumptions 
so far~\cite{Runge:1984,Dreizler:1990lr,Gross:2006,Gross:2012}, which in practice are not always realized. 
In particular \textcite{Runge:1984} showed that for a particular choice of initial conditions
one can introduce a density functional, which in an exact DFT formulation also implies 
the existence of memory terms~\cite{Gross:2006,Gross:2012}. While even for static DFT 
formulations, when memory terms are not necessary, the construction of a density functional 
is still more an art than a science, the density functional for TDDFT should include 
memory terms and the density functional may depend on the initial conditions as well. 
Nevertheless, if the collective motion of the many-body system is ``adiabatic'' one can invoke 
in practical implementations the adiabatic approximation~\cite{Gross:2006,Gross:2012} 
and use a static density functional for a time-dependent problem. This ``adiabatic approximation'' 
is used in all, if not the majority of practical implementations of TDDFT 
and it is the framework adopted in the present study, as also suggested by \textcite{Runge:1984}, 
see their theorem 4, which leads to a natural extension of the Kohn-Sham scheme~\cite{Kohn:1965fk} 
to time-dependent problems.

 \subsection{Upper bounds for the maximum Lyapunov exponents for some
quantum many-body non-relativistic systems}

It is instructive to estimate the upper bound for the Lyapunov
exponent, Eq.~\eqref{eq:lam}, for a realistic system. While there is
plenty of them let us focus on quantum systems at room temperature,
condensed matter systems, low energy nuclear systems, and the unitary
Fermi gas, for which non-relativistic quantum description is
applicable.  

For a condensed matter system at room temperature
\begin{align}
\lambda^\text{CM}_\text{MSS} = \frac{\pi T}{\hbar}  \approx 1.2\times 10^{14}\, \text{s}^{-1}
\end{align}
and combing this with the time it takes for an electron with a
kinetic energy of 3 eV to traverse 100 nm ($\tau_\text{cross} = 1.3
\times 10^{-13}$ s) we obtain
\begin{align}
\lambda^\text{CM}_\text{MSS}\, \tau_\text{cross} \approx 15.
\end{align}
We used 100 nm as the possible size of a quantum billiard, which is
also a reasonable estimate of the mean free path in typical metals.

A characteristic temperature of an excited nucleus is $T={\cal O}(1)$
MeV, which leads to
\begin{align}
\lambda^\text{nuclear}_\text{MSS} \approx  0.016 \,\text{c/fm}= 5.33\times 10^{-26} s^{-1}. \label{eq:Lnuc}
\end{align}
A heavy nucleus has a diameter $\approx 12$ fm, a nucleon has a
velocity $\approx c/4$, and the time it takes to cross a nucleus is
$\tau_\text{cross}\approx 50$ fm/c and thus
\begin{align}
\lambda^\text{nuclear}_\text{MSS} \tau_\text{cross}  ={\cal O}(1).
\end{align} 
The time between two nucleon collisions is not much different $\tau_{coll}
\sim \tau_\text{cross}$.

Another system to which we will turn our attention will be the unitary
Fermi gas (UFG), which is the ``most superfluid system'' known, with a
critical temperature $T_\text{c}\approx 0.17\, \varepsilon_\text{F}$,
where $\varepsilon_\text{F}=\hbar^2k_\text{F}^2/2m$ is the Fermi
energy of the non-interacting Fermi gas with the same number density
$n=k_\text{F}^3/3\pi^2$.  For the UFG the two-body collision time is
$t_\text{coll} \approx \pi\hbar /2\varepsilon_\text{F}$ and thus at
$T_\text{c}$
\begin{align}
\lambda^\text{UFG}_\text{MSS} t_\text{coll} ={\cal O}(1),
\end{align}
thus similar to the estimates we obtained above.

In all these cases the estimates of $\lambda$ given by
Eq.~\eqref{eq:lam} are upper bound estimates and one would expect
according to Eq.~\eqref{equ:equil}
\begin {align}
\lambda_\text{MSS} \tau_\text{equil}\ll 1 
\end{align}
in order to observe chaotic behavior in a quantum system in
statistical equilibrium.  The question arises, whether such strongly
interacting non-relativistic fermions could be chaotic in reality?
The times $\tau_\text{coll}$ and $\tau_{\text{cross}}$ are referred to
as dissipation and scrambling times respectively in
Ref.~\cite{Maldacena:2016} and related studies. What our estimates
here show is that in physical systems of interest in condensed matter,
nuclear, and cold atom physics, there is no separation of scales
between the dissipation and scrambling times and the conjectured onset
of chaoticity time scale $1/\lambda_\text{MSS}$.

\subsection{Time-Dependent Density Functional Theory}

Nevertheless, the initial question we raised is still legitimate: Do
the Schr{\" o}dinger and DFT descriptions, which are identical for one-body
observables, display any chaoticity? This issue which was not
addressed in literature yet, specifically for time-dependent
phenomena, except for a few instances we are aware
of~\cite{Balian:1989,Bulgac:2016x,Bulgac:2019,Shi:2020}. 
In particular  \textcite{Balian:1989} address a narrower question concerning the
Lyapunov stability of the time-dependent Hartree-Fock (TDHF) approximation 
and of the random phase approximation (RPA) only in the case when the 
initial state minimizes either the micro-canonical, canonical, or the grand 
canonical partition function respectively. Since at the minimum of a partition 
function the RPA spectrum is real~\cite{Ring:2004} a Lyapunov instability is 
naturally not expected and formally it follows to be a mathematically correct 
statement~\cite{Balian:1989}, for either a classical or a quantum many-body system.
Our interest here is the more general problem, when the initial state of the
quantum system is a (highly) excited state. 

The DFT does not provide
a recipe to construct the energy density functional and thus any DFT
implementation relies on a number of assumptions.  Even in the case of
electrons in atoms and molecules or in condensed matter systems, even
though the Coulomb interaction is known and the non-relativistic
Schr\"odinger equation is very accurate in principle for
their description. For nuclear systems the situation is even worse,
since the interaction between nucleons is not known with enough
accuracy, and moreover the relativistic effects are not entirely
negligible and the effects of non-nucleon degrees of freedom (virtual
mesons, quarks and gluons) are sometimes required. Nevertheless, as in
the case of Schr{\"o}dinger equation, where we know the form of the
equation and have to provide the interparticle interaction, in case of 
DFT we know the framework and we have to provide the energy 
density functional.
 
In 1964 \textcite{Hohenberg:1964}
proved the remarkable mathematical theorem that the many-electron wave
function of an $N$-electron system in the presence of nuclei is in
one-to-one correspondence with the electron number density. This
result implies that the $N$-electron wave function is fully determined
by the electron number density alone. It also implies that an energy
density functional depending on the electron number density exists,
and its minimum determines the ground-state electron number
density and the ground-state energy of the $N$-electron system, in
full agreement with the solution of the $N$-electron Schr{\"o}dinger
equation. This theorem has been generalized over the years to any
excited state, electron systems in various statistical ensembles, and
time-dependent phenomena~\cite{Dreizler:1990lr,Gross:2006,Gross:2012}.
Since the specific form of the Coulomb electron-electron interaction
plays no role in these proofs, these results apply equally to any
many-fermion systems, irrespective of the nature of the (static) interaction
between them.  One should remember also that DFT does not provide
information about the $n$-body number densities for $n>1$. If the
energy density functional is known then one can extract only the
one-body number density, the energy of the system, and related
observables.

A particular many-fermion system deserves special attention: the
Unitary Fermi Gas (UFG) suggested by
G. F. Bertsch~\cite{Bertsch:1999,Baker:1999,Zwerger:2011}.  In
\textcite{Bertsch:1999} formulation, the UFG is a homogeneous infinite
system of equal number of spin-up and spin-down fermions, interacting
with a zero-range potential and an infinite scattering amplitude. The
UFG energy per particle of a large homogeneous $N$-particle system is
then given by a function
\begin{align}
&\lim_{a\rightarrow \infty}\lim_{r_0\rightarrow 0}
\frac{E(N,V,\hbar, m, r_0, a)}{N}\left |_{\tfrac{N}{V}=\text{const}}   \right .= \xi \frac{3}{5}\varepsilon_F, \label{eq:E}\\
& n = \frac{N}{V}=\frac{k_F^3}{3\pi^2}, \quad \varepsilon_F=\frac{\hbar^2k_F^2}{2m}
\end{align}
where $V, a, r_0, m$ are the volume, scattering length, effective
range, and mass of the fermions, and $n, k_F, \varepsilon_F$ are the
number density, Fermi wave vector, and Fermi energy of the free Fermi
gas with the same $N$ and $V$. For the UFG, apart from mass $m$ and
Planck's constant (which factor out in Eq.~\eqref{eq:E}), the only
dimensional system dependent quantity is the Fermi wave vector. The
dimensionless constant $\xi$ is known as the Bertsch parameter, and by
now has been determined both theoretically
$\xi=0.372(5)$~\cite{Carlson:2011} and experimentally
$\xi=0.376(5)$~\cite{Ku:2011}. From dimensional arguments and
requiring translational, rotational, time-reversal symmetries, the
Galilean invariance, and the renormalizability of the theory one can
show that for an inhomogeneous system the energy density functional of
a UFG is determined as
\begin{align}
\varepsilon({\bm r}) =  \frac{\hbar^2}{m}   \left [  \alpha \frac{\tau({\bm r})}{2} + \beta\frac{ 3(3\pi^2)^{2/3}n^{5/3}({\bm r}) }{5} \right. \\
                                                                \left . + \gamma\frac{ |\nu({\bm r})|^2 }{ n^{1/3}({\bm r}) } - (\alpha-1)\frac{ {\bm j}^2 ( {\bm r}) }{ n({\bm r}) } \right ]+\ldots \nonumber
\end{align}
where we have neglected further gradient corrections like $\sim |\nabla n|^2/n$.
The number, kinetic, anomalous, and current densities are defined as follow
\begin{align} 
n({\bm r})     &= 2\sum_{0<E_k<E_c} |v_k({\bm r})|^2,          \\
\tau({\bm r}) &= 2\sum_{0<E_k<E_c} |{\bm \nabla } v_k({\bm r})|^2, \\
\nu ({\bm r}) &=   \sum_{0<E_k<E_c} u_k({\bm r})v_k^*({\bm r}),      \\
{\bm j}({\bm r}) &=   2\,\textrm{Im} \sum_{0<E_k<E_c}  v_k^*({\bm r}){\bm \nabla} v({\bm r}).
\end{align} 
 They are parametrized in terms of the Bogoliubov quasi-particle wave
functions (qpwfs) $u_k({\bm r}),v_k({\bm r})$, obtained as the
solutions of the self-consistent equations
\begin{align}
\left ( \begin{array}{cc} (h({\bm r})-\mu)   & \Delta({\bm r}) \\ \Delta^*({\bm r}) &  -(h^*({\bm r})-\mu)  \end{array}\right )
&\left ( \begin{array}{r} u_k({\bm r})\\ v_k({\bm r}) \end{array}\right ) \nonumber \\
=E_k&\left ( \begin{array}{r} u_k({\bm r})\\ v_k({\bm r}) \end{array}\right ),\label{eq:slda}
\end{align}
where 
\begin{equation}
h({\bm r})= -\frac{\hbar^2\alpha}{2m}\nabla^2+ \frac{\delta \varepsilon({\bm r})}{\delta n({\bm r})} + V_{\text{ext}}({\bm r}),\quad
\Delta({\bm r})= -\frac{\delta \varepsilon({\bm r})}{\delta \nu^*({\bm r})}.
\end{equation}
Here $V_{\text{ext}}({\bm r})$ stands for a trapping potential and
$\mu$ is the chemical potential.  The need for the kinetic energy
density in addition of the number density was discussed in
Ref.~\cite{Kohn:1965fk}.  The anomalous and current densities are
required to be able to disentangle the superfluid phase from the
normal phase and the static system from the system with flow
respectively. The current density is needed in the case of currents,
e.g. when quantized vortices are present.  The dimensionless
parameters $\alpha, \beta$ and $\gamma$ are extracted from Quantum
Monte Carlo calculations of the homogeneous
UFG~\cite{Bulgac:2007,Bulgac:2011a}.

Many properties of trapped inhomogeneous finite UFG  can
at this point be evaluated using the two different, independent, and
exact  methods, the DFT and the solution of the many-body
Schr\"odinger equation using a Monte Carlo approach (QMC).  The DFT
and QMC results agree to the level of QMC numerical
errors~\cite{Bulgac:2007,Bulgac:2011a}, and therefore the DFT and
Schr{\" o}dinger descriptions indeed agree as expected.

The time-dependent systems are described within the TDDFT framework
\begin{align}
 & i\hbar\begin{pmatrix}
    \dot{u}_k({\bm r},t)\\
    \dot{v}_k({\bm r},t)
  \end{pmatrix} \\
  & = 
  \begin{pmatrix}
    (h({\bm r},t)-\mu)  & \Delta({\bm r},t)          \\ 
    \Delta^*({\bm r},t) &  -(h^*({\bm r},t)-\mu)
  \end{pmatrix}
  \begin{pmatrix}
    u_k({\bm r},t)\\
    v_k({\bm r},t)
  \end{pmatrix}. \nonumber
\end{align}
A direct comparison of the time-dependent Schr{\" o}dinger and TDDFT
solutions has not been yet performed. However, the time-dependent
Schr{\"o}dinger equation for a simplified $N$-fermion superfluid
system, very similar to the
UFG~\cite{Yuzbashyan:2006,Yuzbashyan:2006a,Yuzbashyan:2008} has
similar solutions with the TDDFT for UFG~\cite{Bulgac:2009}. One of
the most spectacular successes of TDDFT was the correct identification
of the generation of vortex rings and their
dynamics~\cite{Bulgac:2011,Bulgac:2014,Wlazlowski:2015,Wlazlowski:2018},
which were initially incorrectly identified as heavy
solitons~\cite{Yefsah:2013} and later confirmed as vortex rings
experimentally as well~\cite{Ku:2014}. Subsequently, it was
demonstrated that the TDDFT for the unitary Fermi gas correctly
describes the evolution of the full solitonic cascade, contrary to
simplified approaches like the standardly applied Gross-Pitaevski
equation~\cite{Wlazlowski:2018}.

We will discuss two other examples of strongly interacting many
fermion systems: nuclear fission and collisions of superfluid
heavy-ions. In the case of nuclear systems we have so far only
approximate forms of the energy density functionals, which are also
significantly more complex than the UFG energy density functional.
In order to describe nuclear systems one has to explicitly add neutron
and proton degrees of freedom and also the spin-orbit interaction and
the corresponding number spin densities, for details see
Refs.~\cite{Stetcu:2011,Stetcu:2014,Bulgac:2016,Wlazlowski:2016,Bulgac:2017,Bulgac:2019b,Bulgac:2020}.
The nuclear qpwfs, separately for neutrons and protons have 
4 components $\{u_k({\bm r},\uparrow), u_k({\bm
r},\downarrow), v_k({\bm r},\uparrow), v_k({\bm r},\downarrow)\}$ to
account for spin 1/2 in the presence of spin-orbit interaction.

Since both the Schr\"{o}dinger equation and the TDDFT lead in
principle to the same number density $n({\bm r},t)$ and related
observables, it is natural to examine the sensitivity to initial
conditions of this quantity.  
We will introduce noise in  the initial state
according to the prescription discussed in Ref.~\cite{Shi:2020}.  
Namely in the case of UFG we will slightly alter the initial values 
of the qpwfs as follows
\begin{align}
  \begin{pmatrix}
    u_k({\bm r},0)\\
    v_k({\bm r},0)
  \end{pmatrix} \rightarrow 
 \begin{pmatrix}
    u_k({\bm r},0)[1+\epsilon \alpha_u({\bm r})]\\
    v_k({\bm r},0)[1+\epsilon \alpha_v({\bm r})]
  \end{pmatrix}, \label{eq:noise}
\end{align}
where $\epsilon$ is the strength of the noise and $\alpha_{u,v}({\bf
r})$ are complex numbers with real and imaginary uniform random
numbers in the interval $[-1, 1]$. The random quantity $\alpha({\bm r})$ 
appear to introduce discontinuities in the modified qpwfs, as neighboring 
lattice points are not correlated. However, this apparent discontinuity 
is at the scale $p_\text{cut}= \hbar \pi/l$, where $l$ is the lattice constant, 
and thus accommodated in our numerical scheme.
In the case of nuclear systems all
four components of the proton and neutron qpwfs are modified
accordingly. The perturbed qpwfs are not orthonormal anymore, but the
solutions of the corresponding TDDFT equations satisfy all expected
symmetries and conservation laws.  According to our discussion in
Section~\ref{sec:qmcm}, this type of perturbation of the initial qpwfs
is equivalent to the change in coordinates and momenta when studying
the sensitivity to initial conditions in the equivalent classical
Hamiltonian framework of either the Schr{\" o}dinger equation or of
the corresponding TDDFT. Since the primary DFT theorem implies that there is a one-to-one
mapping $\Psi\leftrightarrow n$, we can alternatively apply the
perturbation to the densities. The specific method of introducing the
perturbation should not influence the final result.

\begin{figure*}
\includegraphics[width=2\columnwidth]{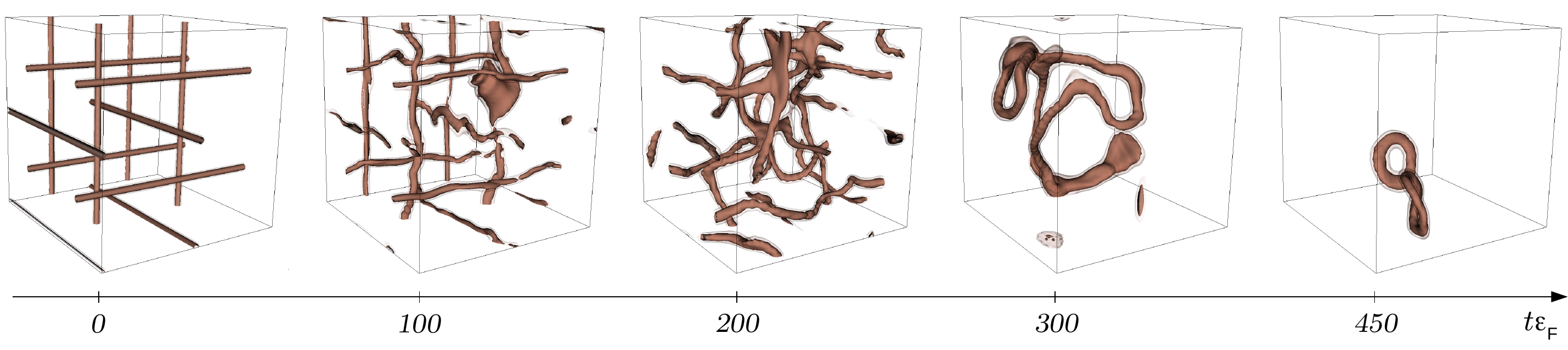}
\caption{ \label{fig:frames} Several consecutive frames demonstrating
the evolution of a system with 12 quantized vortices, perturbed by a
spherical ball stirring them in the time interval $\Delta
t=170\eF^{-1}$, after which the UFG is left to evolve undisturbed. The
system remains superfluid during the entire evolution.  Time is
in units of $1/\varepsilon_F$ and in the case of UFG one also typically 
chooses $\hbar =1$.  }
\end{figure*}

 \begin{figure}
\includegraphics[width=1\columnwidth, trim=75 10 70 0, clip]{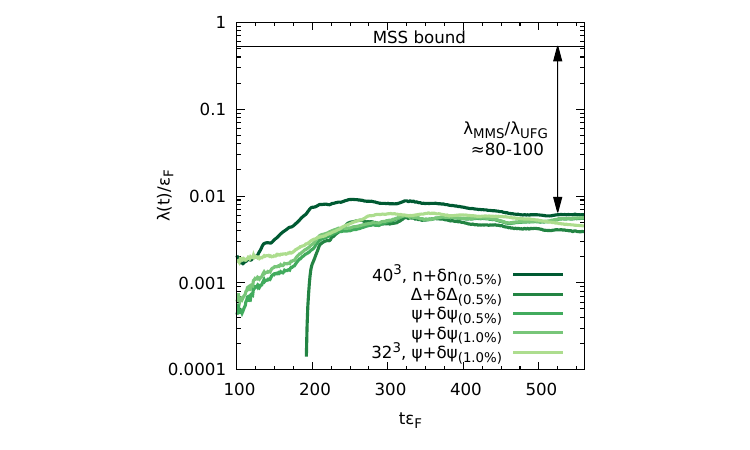}
\caption{ \label{fig:Vortices} The time evolution of $\lambda(t)$, for
different methods of introducing the noise: perturbation of
quasi-particle wave-functions ($\Psi+\delta\Psi$), and initial
hamiltonian through density ($n+\delta n$) or the order parameter
($\Delta+\delta\Delta$). The value after label ($0.5\%$ or $1.0\%$)
indicates strength of the noise $\epsilon$. On top the upper bound by
\textcite{Maldacena:2016} is depicted.  }
\end{figure}

\subsection{Vortex dynamics in a Unitary Fermi gas} 

The first example that we will discuss is the simulation of the
dynamics of 12 quantized vortices in UFG with $N=1004$ fermions in a
simulation box $N=N_xN_yN_z=40^3$, where the dimension of the
equivalent classical phase space is $4\times N^2\approx 1.6
\times10^{10}$.  Initially the vortices are parallel to the sides of
the simulation box and have alternate vorticities, see first frame in
Fig.~\ref{fig:frames}.  At time $t=0$ a spherical ball is 
introduced into the system and starts stirring the system until the time 
$t= 170\eF^{-1}$, when the ball is again  extracted 
(in case of UFG one uses units $m=\hbar=1$) and during this 
time the energy is pumped into the system.  The UFG
is superfluid initially, and the amount of energy pumped into
the system during the time interval $(0, 170)\eF^{-1}$ is sufficiently
small the system remains superfluid. The stirring process destabilizes
the vortex grid and vortices begin to cross and reconnect, as
conjectured by \textcite{Feynman:1955} and in this manner quantum
turbulence (QT) is generated. This type of turbulence is routinely
characterized as a random and chaotic motion of the fluid. Indeed, most
of studies in the QT field are done by means of so called Vortex
Filament Model (VFM), which converts the quantum problem into classical
motion of vortex lines that interact nonlinearly~\cite{VFM2001}.
Since the number of vortices and anti-vortices are equal, eventually
they annihilate each other and after $t\approx 600\eF^{-1}$ there are
no vortices left in the system.

The density plays the central role in the DFT approach, and in order
to measure the Lyapunov exponent we will track this quantity. In
Fig.~\ref{fig:Vortices} we show as a function of time
\begin{align}
\frac{\lambda (t)}{\varepsilon_F} = \frac{1}{2\varepsilon_Ft}\ln 
\frac{ \int \!\!d^3r\, [n_\epsilon({\bm r},t)-n_0({\bm r},t)]^2 }{ \int \!\!d^3r\, [n_\epsilon({\bm r},0)-n_0({\bm r},0)]^2}\label{eq:lv}
\end{align}
where $n_\epsilon({\bm r},t)$ and $n_0({\bm r},t)$ are the densities
arising from perturbed and unperturbed initial states
respectively. These results were obtained using the
code~\cite{W-SLDA}. We used different methods to introduce the initial perturbation.
Noise in the initial qpwfs, in form of Eq.~(\ref{eq:noise}),
of different strengths ($\epsilon=0.5\%$, and $1\%$) provide a very
similar value of $\lambda(t)$ for late times. Alternatively, instead
of perturbing the wave-functions, one can perturb the initial
hamiltonian by adding the noise to the density or to the order
parameter, however, we do not find significant sensitivity of the
final result on the perturbation method. Finally, we have checked that
the result does not depend on the discretization.  Compare the time
series for spatial lattices $40^3$ and $32^3$, where we keep the fixed
volume of the simulation domain and adjusted the lattice spacing
accordingly. All these result clearly demonstrate that for the studied
system the Lyapunov exponent $\lambda_{\textrm{UFG}}=\lim_{t
\rightarrow\infty}\lambda(t)\approx 0.005\eF$.

At times larger than $t\approx600\eF^{-1}$ the vortices disappeared
from the system, the system becomes a uniform gas with many excited
phonons and the time average is a thermal
equilibrium state.  From the energy in the final state, using the
equation of state $E(T)$, which was obtained quite accurately both 
theoretically and experimentally~\cite{Zwerger:2011},  we determine the temperature $T$
corresponding to the equilibrated state, $T/\eF\approx0.17$, which is
practically equal to the critical temperature $T_c$.  The length of simulations 
are however rather short for the system to have thermalized.
At thermal equilibrium at this temperature  the UFG is most likely in the
so-called pseudogap phase, when the long range superfluid correlations
vanished, but a pairing gap is still
present~\cite{Magierski:2009,Magierski:2011,Wlazlowski:2013,Wlazlowski:2013a,Richie-Halford:2020}
and where the ratio of the viscosity to the entropy is
minimal~\cite{Wlazlowski:2015a}, and the system is often referred to as a
perfect fluid.  At this temperature, evaluated using the UFG equation of state~\cite{Bulgac:2006,Bulgac:2008a,Drut:2012} 
\begin{align}
\frac{\lambda_\text{MSS}^\text{UFG}}{\varepsilon_F}=\frac{\pi T}{\varepsilon_F}\approx 0.53, 
\end{align}
which is about two orders of magnitude higher than what we observe in
our calculations.

\begin{figure}
\includegraphics[width=0.9\columnwidth]{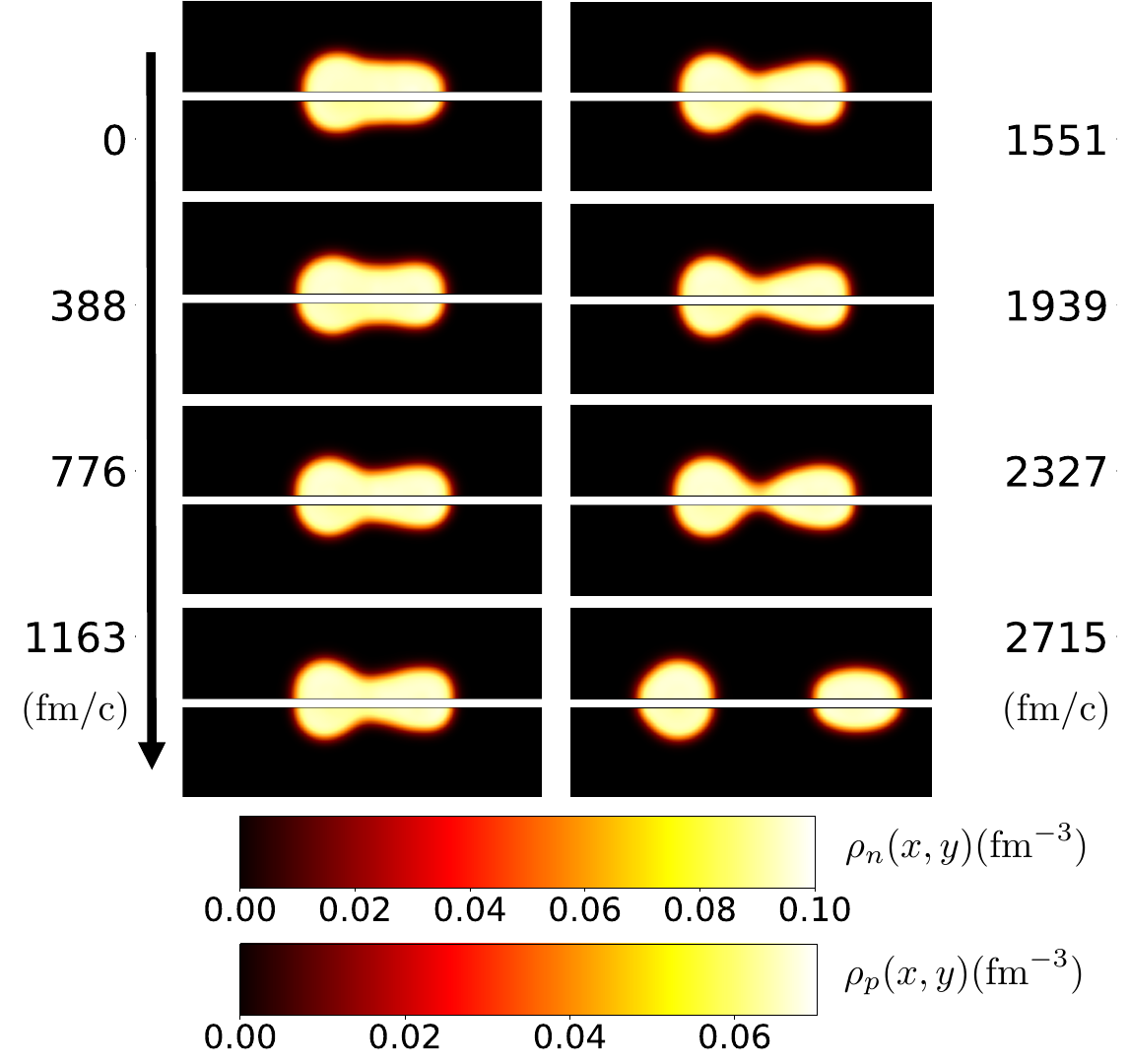}
\includegraphics[width=0.99\columnwidth]{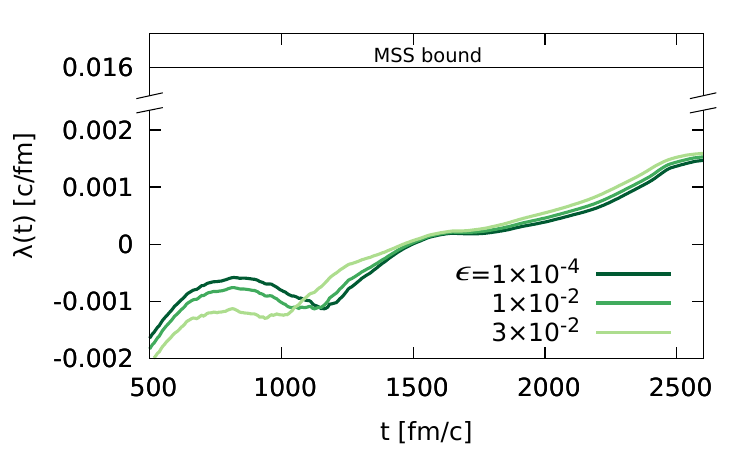}
\caption{ \label{fig:Fission} In the upper panel we display the
evolution of the neutron (upper half of each frame) and proton (lower
half of each frame) number densities of fissioning $^{236}$U in a
simulation box $30^3\times60$ fm$^3$ into a heavy (left) and light (right)
fragments.  The neutron and proton numbers of the heavy (left) and of
the light (right) fragments are ($83.6$, $52.2$) and ($60.4$, $39.8$)
respectively The shapes of the distributions and the neutron and
proton numbers of the fission fragments are hardly affected by the
level of noise level. In the lower panel we show the time evolution of $\lambda(t)$ as obtained from
Eq.~\eqref{eq:lf}.  After scission, which occurs at $t\approx$ 2,400
fm/c, the excited fission fragments emerge, after thermalization,  
with temperatures $T\approx 1$ MeV~\cite{Bulgac:2016,Bulgac:2019}, which results in a $\lambda^\text{nuclear}_\text{MSS}
\approx 0.016$ c/fm.}
\end{figure}

\subsection{Nuclear fission} 

In Ref.~\cite{Shi:2020} we have reported preliminary results on the influence of noise in
initial conditions in TDDFT simulations on some properties of the
emerging nuclear reaction products. In such calculations,  the reaction products
are followed in time only until they are reasonably far spatially separated and they do 
not exchange energy, momentum or particles anymore. At that point the various properties 
of the emerging reaction products are evaluated and that is where we judge the sensitivity
to the initial conditions.  

TDDFT nuclear fission simulations are
equivalent to a classical Hamiltonian system with a dimension of the
phase space $(4\times30^2\times60)^2\approx 4.7\times 10^{10}$ and we
noticed that the maximum Lyapunov exponent is very small or vanishing.
In Fig.~\ref{fig:Fission} we plotted the quantity
\begin{align} \lambda (t) = \frac{1}{2t}\ln \frac{\int \!\!d^3r\,
[n_\epsilon({\bm r},t)-n_0({\bm r},t)]^2}{\int \!\!d^3r\,
[n_\epsilon({\bm r},0)-n_0({\bm r},0)]^2}. \label{eq:lf}
\end{align} It is clear that that the Lyapunov exponent again is very
small, well below the conjectured upper limit $\lambda_\text{MSS}^\text{nuclear}$, 
see Eq.~\eqref{eq:Lnuc}.
Here, we restrict ourselves only to perturbations of initial
quasiparticle wave-functions of the type given in Eq.~(\ref{eq:noise}).  We start the
induced fission simulation relatively late in the evolution of the
reaction $^{235}$U(n,f)~\cite{Hahn:1939}. As it well known for many decades, when a low
energy neutron is absorbed by a nucleus a compound nucleus is
formed~\cite{Bohr:1936,*Bohr:1936a}, which lives for a very long time $\approx
10^{-15}$ sec. before scission. At this excitation energy the level density is very high~\cite{Bethe:1936}
\begin{align}
\rho(E)\propto \exp(\sqrt{aE}), \quad \text{where} \quad a\propto  A,
\end{align}
and $A$ is the atomic number.
During this time the shape of the
compound nucleus changes very slowly from a very compact almost
spherical one to one resembling a peanut at the top of the outer
fission barrier.  During this time the dynamics of the nucleus is
qualitatively similar to a biological cell
division~\cite{Meitner:1939}, in which the nuclear surface tension
competes with the Coulomb interaction, leading ultimately to nuclear
scission~\cite{Meitner:1939}.  

From the top of the outer fission
barrier until scission the nuclear shape evolves relatively fast
$\approx10^{-19}$ sec. This time is however much longer than the  
characteristic single-particle time $\approx 10^{-22}$ sec. Nucleons 
collide with the moving walls of the fissioning nucleus, the nucleus 
heats up, its intrinsic entropy increases, similar to Fermi's acceleration mechanism 
for the origin of cosmic radiation~\cite{Fermi:1949}. The nuclear dynamics is 
highly non-equilibrium~\cite{Bulgac:2016,Bulgac:2019b,Bulgac:2020} and the collective
or the nuclear shape motion of the fissioning nuclear system 
is overdamped.

The dynamics of the fissioning nucleus from the outer fission barrier
until scission is similar to some extent to the evolution of a complex
molecule, which undergoes some big rearrangements of its atoms. The
total energy is conserved, but energy is continuously converted from
the nuclear configurational energy in the molecule into the energy of
the electron cloud. In the case of the fissioning nucleus, the
configuration energy is approximately equal to the sum of the Coulomb
energy and the surface energy. While the nucleus evolves from the
outer fission barrier towards the scission configuration it elongates,
while preserving its volume. The surface area increases and the
surface energy increases proportionally, but at the same time the
average separation of the electric charges increases and the Coulomb
energy decreases. At small elongations of the fissioning nucleus the
surface potential energy dominates, but as soon as the nucleus reaches
the outer fission barrier the sum of surface and Coulomb energies is
dominated by the Coulomb energy~\cite{Meitner:1939}.  The approximate
sum of the surface potential and Coulomb energy decreases by $\approx
15-20$ MeV from the top of the fission barrier until scission.  This
energy difference is converted into internal energy of the
system. Only a small part ($\approx 10\%$) of the energy
is converted into kinetic flow energy of the nuclear fluid, 
the rest being converted into  heat.

The other noticeable aspect of the
fission dynamics from the outer fission barrier until scission put in
evidence in TDDFT simulations is that during the descent from the
neighborhood of the outer fission barrier is that the shape dynamics
has a focusing character. Since the shape evolution is
overdamped the collective motion is similar to that of a parachute,
which follows the steepest descent with a rather small ``terminal
velocity.''  As a result, trajectories with rather distinct initial conditions 
behave like a group of parachutists, who jump out of a plane at
different times and they land relatively close to each other, as both their
vertical and longitudinal velocities are damped. This is the reason
why the $\lambda(t)$ in the case of fission is negative before the
scission occurs. 

Note that after the scission, the number of nucleons in each fragment,
as well as their energies, are fixed.  Since these are measured
observables experimentally, this moment defines for all practical purposes 
the meaning of the ``infinite-time limit.''  In spite of the complex nature of the problem, the
evolution turns out to be very weakly sensitive to the initial noise,
with the corresponding Lyapunov exponent being well below the
prediction of \textcite{Maldacena:2016}.

\begin{figure}
\includegraphics[width=0.9\columnwidth]{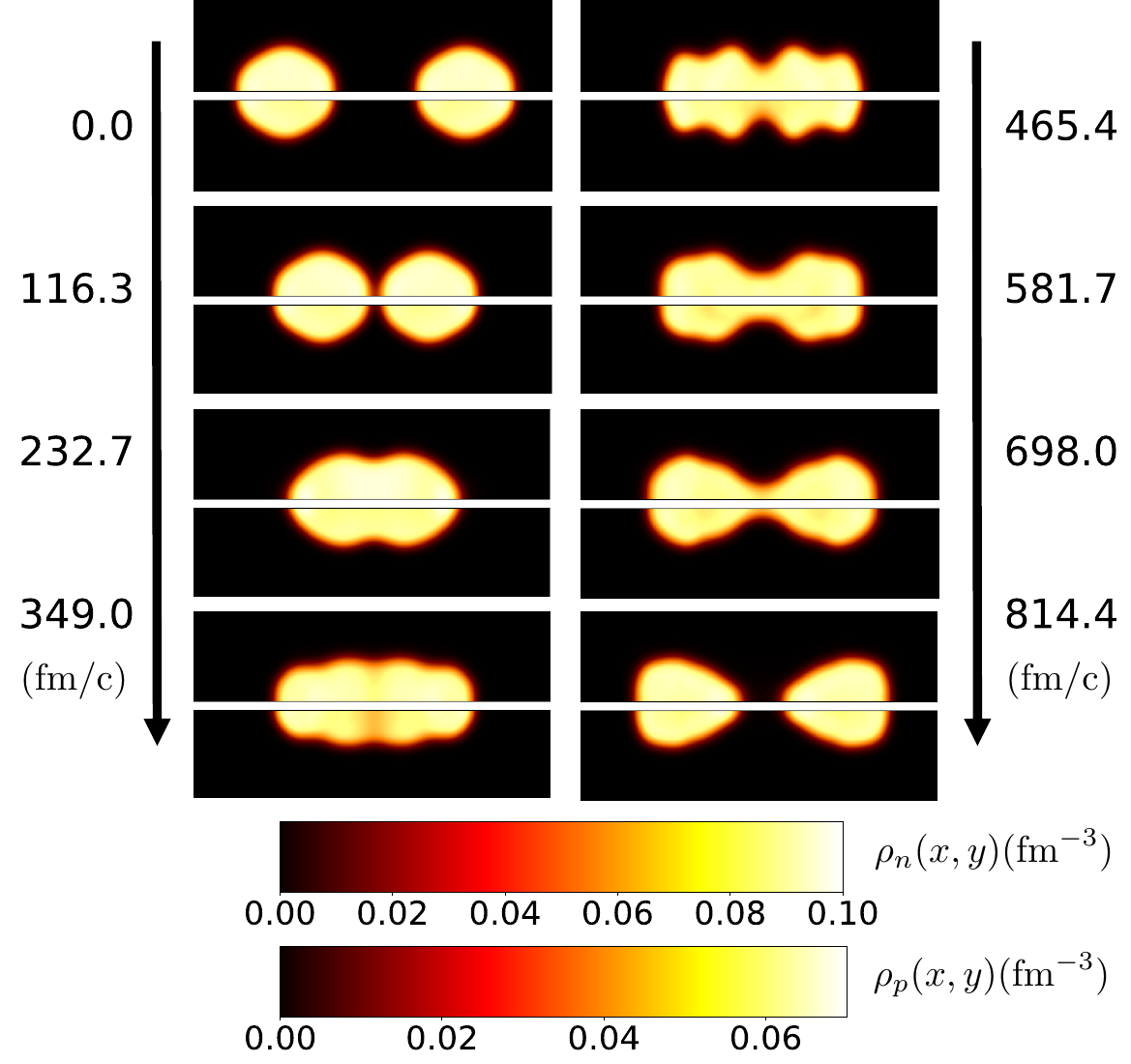}
\includegraphics[width=0.99\columnwidth]{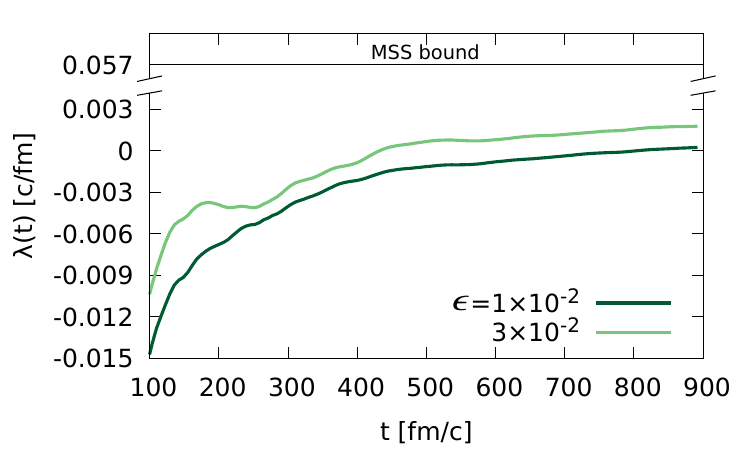}
\caption{ \label{fig:Collision} In the upper panel we display the
evolution of the neutron (upper half of each frame) and proton (lower
half of each frame) number densities of fissioning $^{238}$U+$^{238}$U
collision in a simulation box $30^3\times64$ fm$^3$ into two highly
excited and not yet equilibrated fragments.  The shapes of the neutron
and proton distributions are hardly affected by the noise level. In the lower panel, the
time evolution of $\lambda(t)$, Eq.~\eqref{eq:lf}, in the case
collision of two heavy-ions $^{238}$U+$^{238}$U at a center-of-mass
energy of 1,200 MeV is shown.  After separation the excited fragments emerge,
after thermalization,
with temperatures $T=3.55$ MeV, which results in a
$\lambda^\text{nuclear}_\text{MSS} \approx 0.057$ c/fm.}
\end{figure}

\subsection{Collisions of heavy-ions}
A second nuclear physics example is from heavy-ion collisions
$^{238}$U+$^{238}$U at a center-of-mass energy of 1,200 MeV.  In this
case the dimension of the equivalent phase space is
$(4\times30^3\times64)^2\approx 5.3\times10^{10}$.  After the collision
the two fragments are highly excited as the final total kinetic energy is
only $\approx 600$ MeV. By changing the initial kinetic energy from 600 MeV to 1,600 MeV
the final kinetic energy barely changes, and the final estimated temperatures of the fragments 
varies by at most a factor of 2. The case illustrated in Fig.~\ref{fig:Collision} 
is in the middle of this energy range, 
see also Ref.~\cite{Shi:2020}. From the excitation energy of the reaction 
fragments one can extract their equilibrium temperatures $T\approx 3.5$ MeV, 
obtained using Bethe formula~\cite{Bethe:1936,Shi:2020},
which according to Eq.~\eqref{eq:lam} leads to an upper limit $\lambda_\text{MSS}
= 0.057$ c/fm.

Upon adding noise at the level $\epsilon = 0.03$, see
Eq.~\eqref{eq:noise}, the nuclei acquire a total initial excitation
energy of about 50 MeV, corresponding to an initial temperature
$T\approx 1.5$ MeV.  In a time $\approx 250$ fm/c the two nuclei touch
``noses'' and remain in contact up to a time $\approx$ 700 fm/c.
Until the nuclei coalesce, which a relatively fast process, only the
long-range Coulomb interaction between them leads a small charge
polarization of the colliding nuclei.  The two nuclei form a
di-nucleus at about 300 fm/c, which starts separating into two
fragments around 550 fm/c and the contact time is thus relatively
short.  Similarly to the fission process, after the separation,
fragments and their properties are fixed, and there is no reason to
consider the evolution in the context of extracting of the Lyapunov
exponent. Nevertheless, the $\lambda(t)$ appears to be almost
saturated, and again by about two orders of magnitude smaller than the
upper bound.

\section{Conclusions} 

We have analyzed several strongly interacting quantum many-fermion
systems, in which superfluid correlations are very important
in order to correctly describe their non-equilibrium dynamics, with a
number of degrees of freedom of the order of $10^{10}$ or
larger. Their dynamical evolution is described within an extension of
the TDDFT to superfluid systems, which is expected to be
mathematically equivalent to the description of the same systems using
the time-dependent many-body Schr{\"o}dinger equation, but only at the
level of the one-body densities.  Since the Schr\"odinger equation is
linear no exponential sensitivity to the variations of the initial
conditions exists.  Therefore, if one replaces the initial many-body
wave function $\Psi(x_1,...,x_N,t)
\rightarrow\Psi(x_1,...,x_N,0)+\delta \Psi(x_1,...,x_N,0)$
\begin{align} 
\langle \Psi(0)|\delta \Psi(0)\rangle \equiv  \langle \Psi(t)|\delta \Psi(t)\rangle\ll 1 
\end{align}
one does not expect an exponential divergence of the two wave
functions. The overlap of two arbitrary wave functions $\langle
\Psi_1(0)|\Psi_2(0)\rangle = \langle \Psi_1(t)|\Psi_2(t)\rangle$ is
always preserved in the case of a time-dependent Schr{\" o}dinger
equation. As we mentioned in the introduction, the Schr\"odinger
equation is also mathematically equivalent to a system of classical
coupled harmonic oscillators, for which the absence of chaos is well
known.

Even though the Schr\"odinger description is in principle
mathematically identical to the DFT description at the one-body
density level, the DFT equations are non-linear, their sensitivity, or
the absence of such sensitivity, to the initial conditions has never
been demonstrated. In this study we have presented strong arguments
that for very large realistic quantum many-body systems the Lyapunov
exponents within a DFT description are much smaller than the
conjectured by \textcite{Maldacena:2016} upper limit. However, even
though the Lyapunov exponents are non-vanishing, they are small enough
as not to alter the quality of the conclusions inferred within the
present TDDFT framework.

We presented typical simulation results for three different 
strongly interacting many-fermion superfluid systems. By changing 
various  parameters, particle number, energy density functionals, 
energy of the system, and so forth, the results remain qualitatively similar.  
In all cases the Lyapunov exponents are smaller by a
factor of $10\ldots 100$ than the upper limit conjectured by
\textcite{Maldacena:2016,Tsuji:2018}, but likely not vanishing as one
would have expected, if the TDDFT is mathematically equivalent to the
Schr{\"o}dinger equation at the level of one-body number density. The
most obvious offender appears to be the energy density functional
used. Strictly speaking, in the TDDFT the evolution equations are
expected to have memory terms~\cite{Gross:2006,Gross:2012}, which are
absent in all the cases we have discussed here. If the quantum
evolution is ``slow'' one can invoke the adiabatic limit of the energy
density functional, when the memory terms can be suceptibly neglected, which was
assumed in all the cases we studied.  Thus the measure to what extent
the Lyapunov exponents are not vanishing points to the accuracy to the
current implementation of the TDDFT formalism in the case of strongly
interacting fermion systems.

A second cause leading to non-vanishing Lyapunov exponents, which we
cannot fully rule out at this time, could be due to the accumulation
of numerical errors in evaluating the evolution of a system with
${\cal O}(10^{10}) $ phase space variables in current numerical
implementations.  The test we have performed so far indicate that our
codes are however numerically quite accurate~\cite{Shi:2020,W-SLDA}.  On the
other hand, the theory is operative only if one can use it to make
predictions. It is hard to achieve it without practical
implementation. The Nobel Committee already recognized this critical
aspect, when awarding the prize in 1998 to Walter Kohn ``for his
development of the density-functional theory.'' 
Thus, in reality, one
should measure the quality of the TDDFT by taking into account
contributions to the Lyapunov exponent coming from both sources:
the accuracy of the underlying energy density functional and its numerical
realization.

The absence of positive Lyapunov exponents 
within TDDFT is crucial when one considers extracting 
the magnitude of fluctuations of various physical quantities within a mean 
field approach~\cite{Balian:1984,Balian:1984a,Balian:1985,Simenel:2011,Simenel:2012,Scamps:2015,Williams:2018}, 
such a neutron, proton, and neutron-proton particle variances. \textcite{Balian:1984} 
designed a variational approach based on an action-like functional
\begin{align}
{\cal I} &= \Tr \, {\cal A}(t_1){\cal D}(t_1) \nonumber \\
&-\int_{t_0}^{t_1}\!\!\!dt \,\Tr \left ({\cal A}(t)\frac{d {\cal D}(t)}{dt} + i[H,{\cal D}(t)]\right ) ,
\end{align} 
where ${\cal D}(t)$ and ${\cal A}(t)$ are two time-dependent operators of the same nature as 
a density operator and an observable, and $H$ is the Hamiltonian. Unfortunately 
this action-like functional cannot be minimized and  can only be made stationary with 
respect to arbitrary variations of both ${\cal D}(t)$ and ${\cal A}(t)$,
subject to boundary conditions
\begin{align}
{\cal D}(t_0) = D, \quad {\cal A}(t_1)= A,
\end{align}  
where $D$ is the initial value of the density matrix and $A$ the final value 
of the observable of interest. This approach appears to be useful in practice 
in carefully chosen situations. The dispersion of the observable ${\cal A}(t_1)$
can then be determined from the limit
\begin{align}
\langle \langle {\cal A}^2(t_1)\rangle \rangle  = \frac{1}{2} \lim_{\epsilon\rightarrow 0} 
\frac{\langle  [ D -\sigma (t_0,\epsilon)]^2 \rangle }{\epsilon^2},
\end{align}
where
\begin{align}
\sigma (t_1,\epsilon) = {\cal A}(t_1,\epsilon) = e^{i\epsilon A}{\cal D}(t_1) e^{-i\epsilon A},
\end{align}
where one chooses in practice $\epsilon = 0.001\ldots 0.0001$. One thus 
needs to propagate  the equations of motion forward in time for ${\cal D}(t)$ 
and backward in time for $\sigma(t,\epsilon)$ , and in the presence of chaoticity 
such results will become totally unpredictable. As we have 
shown here and partially in Ref.~\cite{Shi:2020} the TDDFT equations lack sensitivity 
to initial conditions with high numerical accuracy. 

For practical reasons, the resilience of the TDDFT framework to
describe the evolution of dense quantum many-body systems with respect
to the level of noise is remarkable. On the other hand, the present
results show a different aspect of the decades old question: Under
what conditions can we observe the sensitivity to initial conditions
in realistic time-dependent non-equilibrium phenomena in dense quantum
many-body systems in the absence of the limit $\hbar \rightarrow 0$?
In TDDFT applications it makes sense to examine the validity of the
theory up to the point in time when predictions are expected. In the
case of quantum turbulence at the stage when the quantized vortices
have already decayed, in the case of nuclear fission and heavy-ion
collisions in nuclear physics that is at the time when the properties of the
emerging fragments have been defined.  These time scales are to some
extent not very long, but large enough to allow for the evidence of
sensitivity to initial conditions. Niels Bohr's compound
nucleus~\cite{Bohr:1936,*Bohr:1936a}, formed for example in the case of the
neutron induced fission of $^{235}$U(n,f)~\cite{Hahn:1939}, is an
example of a quantum many-fermion system which evolves for extremely
long times $\tau \approx 10^{-15}$ sec.  In order to have a noticeable
effect of the sensitivity to initial conditions in a compound nucleus
one would have to distinguish or measure properties of different
quantum states, characterized by different quantum numbers,
thus be sensitive to energy differences comparable to 
the average level separation $\Delta E = 1/\rho(E)$.
One could argue also that one would have to discriminate
between systems with spectral properties characterized either of the
random matrix theory - typically the Gaussian orthogonal  
ensemble (GOE) -  or of the integrable or Poisson 
type~\cite{Bohigas:1984,Balian:1974,Berry:1981,Heller:1984,Gutzwiller:1977,Larkin:1969,Mehta:1991,Deutsch:1991,Srednicki:1994,Volya:2020},
in order to distinguish between various final configurations, thus be
able to probe energy differences larger than the separation between
several levels in complex quantum systems at significant excitation energies,
where the level density is quite  high.  Induced fission~\cite{Hahn:1939}, one of the three 
cases we considered here, is an example where 
a compound nucleus is formed~\cite{Bohr:1936,*Bohr:1936a}, at an excitation energy with very high level
density~\cite{Bethe:1936} and the outcomes of the reaction are largely independent 
of the excitation mechanism. Such long time scales are likely
unachievable within an unrestricted real-time quantum framework and
using present and near future computational resources.
 
\begin{acknowledgments}

AB thanks V. Zelevinsky for reading the initial draft of the manuscript and for a
number of comments.  AB was supported by U.S. Department of Energy,
Office of Science, Grant No. DE-FG02-97ER41014 and in part by NNSA
cooperative Agreement DE-NA0003841.  The work of IA is based upon work
supported by the Department of Energy, National Nuclear Security
Administration, under Award Number DE-NA0003841.  The work of GW was
supported by the Polish National Science Center (NCN) under Contracts
No. UMO-2017/26/E/ST3/00428.  This research used resources of the Oak
Ridge Leadership Computing Facility, which is a U.S. DOE Office of
Science User Facility supported under Contract No. DE-AC05-00OR22725
and of the National Energy Research Scientific computing Center, which
is supported by the Office of Science of the U.S. Department of Energy
under Contract No. DE-AC02-05CH11231.
\end{acknowledgments}  

\providecommand{\selectlanguage}[1]{}
\renewcommand{\selectlanguage}[1]{}

\bibliography{latest_fission}

\begin{thebibliography}{92}%
\makeatletter
\providecommand \@ifxundefined [1]{%
 \@ifx{#1\undefined}
}%
\providecommand \@ifnum [1]{%
 \ifnum #1\expandafter \@firstoftwo
 \else \expandafter \@secondoftwo
 \fi
}%
\providecommand \@ifx [1]{%
 \ifx #1\expandafter \@firstoftwo
 \else \expandafter \@secondoftwo
 \fi
}%
\providecommand \natexlab [1]{#1}%
\providecommand \enquote  [1]{``#1''}%
\providecommand \bibnamefont  [1]{#1}%
\providecommand \bibfnamefont [1]{#1}%
\providecommand \citenamefont [1]{#1}%
\providecommand \href@noop [0]{\@secondoftwo}%
\providecommand \href [0]{\begingroup \@sanitize@url \@href}%
\providecommand \@href[1]{\@@startlink{#1}\@@href}%
\providecommand \@@href[1]{\endgroup#1\@@endlink}%
\providecommand \@sanitize@url [0]{\catcode `\\12\catcode `\$12\catcode
  `\&12\catcode `\#12\catcode `\^12\catcode `\_12\catcode `\%12\relax}%
\providecommand \@@startlink[1]{}%
\providecommand \@@endlink[0]{}%
\providecommand \url  [0]{\begingroup\@sanitize@url \@url }%
\providecommand \@url [1]{\endgroup\@href {#1}{\urlprefix }}%
\providecommand \urlprefix  [0]{URL }%
\providecommand \Eprint [0]{\href }%
\providecommand \doibase [0]{http://dx.doi.org/}%
\providecommand \selectlanguage [0]{\@gobble}%
\providecommand \bibinfo  [0]{\@secondoftwo}%
\providecommand \bibfield  [0]{\@secondoftwo}%
\providecommand \translation [1]{[#1]}%
\providecommand \BibitemOpen [0]{}%
\providecommand \bibitemStop [0]{}%
\providecommand \bibitemNoStop [0]{.\EOS\space}%
\providecommand \EOS [0]{\spacefactor3000\relax}%
\providecommand \BibitemShut  [1]{\csname bibitem#1\endcsname}%
\let\auto@bib@innerbib\@empty
\bibitem [{\citenamefont {Poincar\'e}(1890)}]{Poincare:1890}%
  \BibitemOpen
  \bibfield  {author} {\bibinfo {author} {\bibfnamefont {H.}~\bibnamefont
  {Poincar\'e}},\ }\bibfield  {title} {\enquote {\bibinfo {title} {{Sur le
  probl\`eme de trois corps et les \'equation de la dynamique}},}\ }\href@noop
  {} {\bibfield  {journal} {\bibinfo  {journal} {Acta Mathematica}\ }\textbf
  {\bibinfo {volume} {13}},\ \bibinfo {pages} {1--270} (\bibinfo {year}
  {1890})}\BibitemShut {NoStop}%
\bibitem [{\citenamefont {R\"agstedt}()}]{mittag}%
  \BibitemOpen
  \bibfield  {author} {\bibinfo {author} {\bibfnamefont {M.}~\bibnamefont
  {R\"agstedt}},\ }\href@noop {} {\enquote {\bibinfo {title} {{From order to
  chaos: the prize competition in honor of King Oscar II, \\
  http://www.mittag-leffler.se/library/henri-poincare}},}\ }\BibitemShut
  {NoStop}%
\bibitem [{\citenamefont {Lyapunov}(1992)}]{Lyapunov:1992}%
  \BibitemOpen
  \bibfield  {author} {\bibinfo {author} {\bibfnamefont {A.~M.}\ \bibnamefont
  {Lyapunov}},\ }\bibfield  {title} {\enquote {\bibinfo {title} {The general
  problem of the stability of motion},}\ }\href {\doibase
  10.1080/00207179208934253} {\bibfield  {journal} {\bibinfo  {journal}
  {International Journal of Control}\ }\textbf {\bibinfo {volume} {55}},\
  \bibinfo {pages} {531} (\bibinfo {year} {1992})}\BibitemShut {NoStop}%
\bibitem [{\citenamefont {Lorenz}(1963)}]{Lorenz:1963}%
  \BibitemOpen
  \bibfield  {author} {\bibinfo {author} {\bibfnamefont {E.~N.}\ \bibnamefont
  {Lorenz}},\ }\bibfield  {title} {\enquote {\bibinfo {title} {{Deterministic
  Nonperiodic Flow}},}\ }\href {\doibase
  10.1175/1520-469(1963)020<0130:DNF>2.0.CO;2} {\bibfield  {journal} {\bibinfo
  {journal} {Journal of the Atmospheric Sciences}\ }\textbf {\bibinfo {volume}
  {20}},\ \bibinfo {pages} {130} (\bibinfo {year} {1963})}\BibitemShut
  {NoStop}%
\bibitem [{\citenamefont {Kol}(2021)}]{Kol:2021}%
  \BibitemOpen
  \bibfield  {author} {\bibinfo {author} {\bibfnamefont {B.}~\bibnamefont
  {Kol}},\ }\bibfield  {title} {\enquote {\bibinfo {title} {Flux-based
  statistical prediction of three-body outcomes},}\ }\href {\doibase
  10.1007/s10569-021-10015x} {\bibfield  {journal} {\bibinfo  {journal}
  {{Celestial Mechanics and Dynamical Astronomy}}\ }\textbf {\bibinfo {volume}
  {133}},\ \bibinfo {pages} {17} (\bibinfo {year} {2021})}\BibitemShut
  {NoStop}%
\bibitem [{\citenamefont {Manvadkar}\ \emph {et~al.}(2020)\citenamefont
  {Manvadkar}, \citenamefont {Trani},\ and\ \citenamefont
  {Leigh}}]{Manwadkar:2020}%
  \BibitemOpen
  \bibfield  {author} {\bibinfo {author} {\bibfnamefont {V.}~\bibnamefont
  {Manvadkar}}, \bibinfo {author} {\bibfnamefont {A.~A.}\ \bibnamefont
  {Trani}}, \ and\ \bibinfo {author} {\bibfnamefont {N.~W.~C.}\ \bibnamefont
  {Leigh}},\ }\bibfield  {title} {\enquote {\bibinfo {title} {{Chaos and L\'evy
  lights in the three-body problem}},}\ }\href {\doibase
  10.1093/mnras/staa1722} {\bibfield  {journal} {\bibinfo  {journal} {MNRAS}\
  }\textbf {\bibinfo {volume} {497}},\ \bibinfo {pages} {3694} (\bibinfo {year}
  {2020})}\BibitemShut {NoStop}%
\bibitem [{\citenamefont {Manwadkar}\ \emph {et~al.}(2021)\citenamefont
  {Manwadkar}, \citenamefont {Kol}, \citenamefont {Trani},\ and\ \citenamefont
  {Leigh}}]{Manwadkar:2021}%
  \BibitemOpen
  \bibfield  {author} {\bibinfo {author} {\bibfnamefont {V.}~\bibnamefont
  {Manwadkar}}, \bibinfo {author} {\bibfnamefont {B.}~\bibnamefont {Kol}},
  \bibinfo {author} {\bibfnamefont {A.~A.}\ \bibnamefont {Trani}}, \ and\
  \bibinfo {author} {\bibfnamefont {N.~W.~C.}\ \bibnamefont {Leigh}},\
  }\bibfield  {title} {\enquote {\bibinfo {title} {{Testing thre Flux-based
  statisitical prediction of the Three-Body Problem}},}\ }\href {\doibase
  10.1093/mnras/stab1689} {\bibfield  {journal} {\bibinfo  {journal} {MNRAS}\
  }\textbf {\bibinfo {volume} {506}},\ \bibinfo {pages} {692} (\bibinfo {year}
  {2021})}\BibitemShut {NoStop}%
\bibitem [{\citenamefont {Ginat}\ and\ \citenamefont
  {Perets}(2021)}]{Ginat:2021}%
  \BibitemOpen
  \bibfield  {author} {\bibinfo {author} {\bibfnamefont {Y.~B.}\ \bibnamefont
  {Ginat}}\ and\ \bibinfo {author} {\bibfnamefont {H.~B.}\ \bibnamefont
  {Perets}},\ }\bibfield  {title} {\enquote {\bibinfo {title} {Analytical,
  statistical approximate solution of dissipative and nondissipative
  binary-single stellar encounters},}\ }\href {\doibase
  10.1103/PhysRevX.11.031020} {\bibfield  {journal} {\bibinfo  {journal} {Phys.
  Rev. X}\ }\textbf {\bibinfo {volume} {11}},\ \bibinfo {pages} {031020}
  (\bibinfo {year} {2021})}\BibitemShut {NoStop}%
\bibitem [{\citenamefont {Pesin}(1977)}]{Pesin:1977}%
  \BibitemOpen
  \bibfield  {author} {\bibinfo {author} {\bibfnamefont {Y.~B.}\ \bibnamefont
  {Pesin}},\ }\bibfield  {title} {\enquote {\bibinfo {title} {{Characteristic
  Lyapunov Exponents and Smooth Ergodic Theory}},}\ }\href {\doibase
  10.1070/RM1977v032n04ABEH001693} {\bibfield  {journal} {\bibinfo  {journal}
  {Russian Math. Surveys}\ }\textbf {\bibinfo {volume} {32}},\ \bibinfo {pages}
  {55} (\bibinfo {year} {1977})}\BibitemShut {NoStop}%
\bibitem [{\citenamefont {Kaplan}\ and\ \citenamefont
  {Yorke}(1979)}]{Kaplan:1979}%
  \BibitemOpen
  \bibfield  {author} {\bibinfo {author} {\bibfnamefont {J.}~\bibnamefont
  {Kaplan}}\ and\ \bibinfo {author} {\bibfnamefont {J.}~\bibnamefont {Yorke}},\
  }\href@noop {} {\emph {\bibinfo {title} {Chaotic behavior of multidimensional
  difference equations}}},\ edited by\ \bibinfo {editor} {\bibfnamefont
  {H.~O.}\ \bibnamefont {Peitgen}}\ and\ \bibinfo {editor} {\bibfnamefont
  {H.~O.}\ \bibnamefont {Walther}},\ Vol.\ \bibinfo {volume} {{Functional
  Differential Equations and Approximations of Fixed Points}}\ (\bibinfo
  {publisher} {Springer, New York},\ \bibinfo {year} {1979})\BibitemShut
  {NoStop}%
\bibitem [{\citenamefont {Landau}\ and\ \citenamefont
  {Lifshitz}(1966)}]{LL6:1966}%
  \BibitemOpen
  \bibfield  {author} {\bibinfo {author} {\bibfnamefont {L.~D.}\ \bibnamefont
  {Landau}}\ and\ \bibinfo {author} {\bibfnamefont {E.~M.}\ \bibnamefont
  {Lifshitz}},\ }\href@noop {} {\emph {\bibinfo {title} {Fluid Mechanics}}},\
  \bibinfo {series} {Course of Theoretical Physics}, Vol.~\bibinfo {volume}
  {6}\ (\bibinfo  {publisher} {Butterworth-Heinemann},\ \bibinfo {address}
  {Oxford},\ \bibinfo {year} {1966})\BibitemShut {NoStop}%
\bibitem [{\citenamefont {Mehta}(1991)}]{Mehta:1991}%
  \BibitemOpen
  \bibfield  {author} {\bibinfo {author} {\bibfnamefont {M.~L.}\ \bibnamefont
  {Mehta}},\ }\href@noop {} {\emph {\bibinfo {title} {{Random Matrices}}}}\
  (\bibinfo  {publisher} {Academic Press Inc., Boston},\ \bibinfo {year}
  {1991})\BibitemShut {NoStop}%
\bibitem [{\citenamefont {Bohigas}\ \emph {et~al.}(1984)\citenamefont
  {Bohigas}, \citenamefont {Giannoni},\ and\ \citenamefont
  {Schmit}}]{Bohigas:1984}%
  \BibitemOpen
  \bibfield  {author} {\bibinfo {author} {\bibfnamefont {O.}~\bibnamefont
  {Bohigas}}, \bibinfo {author} {\bibfnamefont {M.~J.}\ \bibnamefont
  {Giannoni}}, \ and\ \bibinfo {author} {\bibfnamefont {C.}~\bibnamefont
  {Schmit}},\ }\bibfield  {title} {\enquote {\bibinfo {title}
  {{Characterization of Chaotic Quantum Spectra and Universality of Level
  Fluctuation Laws}},}\ }\href {\doibase 10.1103/PhysRevLett.52.1} {\bibfield
  {journal} {\bibinfo  {journal} {Phys. Rev. Lett.}\ }\textbf {\bibinfo
  {volume} {52}},\ \bibinfo {pages} {1} (\bibinfo {year} {1984})}\BibitemShut
  {NoStop}%
\bibitem [{\citenamefont {Balian}\ and\ \citenamefont
  {Bloch}(1974)}]{Balian:1974}%
  \BibitemOpen
  \bibfield  {author} {\bibinfo {author} {\bibfnamefont {R.}~\bibnamefont
  {Balian}}\ and\ \bibinfo {author} {\bibfnamefont {C.}~\bibnamefont {Bloch}},\
  }\bibfield  {title} {\enquote {\bibinfo {title} {{Solution of the
  Schr{\"o}dinger equation in terms of classical path}},}\ }\href {\doibase
  10.1016/0003-4916(74)90421-7} {\bibfield  {journal} {\bibinfo  {journal}
  {Ann. Phys.}\ }\textbf {\bibinfo {volume} {85}},\ \bibinfo {pages} {514}
  (\bibinfo {year} {1974})}\BibitemShut {NoStop}%
\bibitem [{\citenamefont {Berry}\ and\ \citenamefont
  {Tabor}(1976)}]{Berry:1976}%
  \BibitemOpen
  \bibfield  {author} {\bibinfo {author} {\bibfnamefont {M.~V.}\ \bibnamefont
  {Berry}}\ and\ \bibinfo {author} {\bibfnamefont {M.}~\bibnamefont {Tabor}},\
  }\bibfield  {title} {\enquote {\bibinfo {title} {{Closed Orbits and the
  Regular Bound Spectrum}},}\ }\href {\doibase 10.2307/79042} {\bibfield
  {journal} {\bibinfo  {journal} {Proc. Royal Society of London. Series A,}\
  }\textbf {\bibinfo {volume} {349}},\ \bibinfo {pages} {101} (\bibinfo {year}
  {1976})}\BibitemShut {NoStop}%
\bibitem [{\citenamefont {Gutzwiller}(1977)}]{Gutzwiller:1977}%
  \BibitemOpen
  \bibfield  {author} {\bibinfo {author} {\bibfnamefont {M.~C.}\ \bibnamefont
  {Gutzwiller}},\ }\bibfield  {title} {\enquote {\bibinfo {title} {{Periodic
  Orbits and Classical Quantization Condition}},}\ }\href {\doibase
  10.1063/1.1665596} {\bibfield  {journal} {\bibinfo  {journal} {J. Math.
  Phys}\ }\textbf {\bibinfo {volume} {12}},\ \bibinfo {pages} {343} (\bibinfo
  {year} {1977})}\BibitemShut {NoStop}%
\bibitem [{\citenamefont {Berry}(1981)}]{Berry:1981}%
  \BibitemOpen
  \bibfield  {author} {\bibinfo {author} {\bibfnamefont {M.~V.}\ \bibnamefont
  {Berry}},\ }\bibfield  {title} {\enquote {\bibinfo {title} {{Quantizing a
  classically ergodic system: Sinai's billiard and the KKR method},},}\ }\href
  {\doibase 10.1016/0003-4916(81)90189-5} {\bibfield  {journal} {\bibinfo
  {journal} {Ann. Phys.}\ }\textbf {\bibinfo {volume} {131}},\ \bibinfo {pages}
  {163} (\bibinfo {year} {1981})}\BibitemShut {NoStop}%
\bibitem [{\citenamefont {Heller}(1984)}]{Heller:1984}%
  \BibitemOpen
  \bibfield  {author} {\bibinfo {author} {\bibfnamefont {E.~J.}\ \bibnamefont
  {Heller}},\ }\bibfield  {title} {\enquote {\bibinfo {title} {{Bound-State
  Eigenfunctions of Classically Chaotic Hamiltonian Systems: Scars of Periodic
  Orbits}},}\ }\href {\doibase 10.1103/PhysRevLett.53.1515} {\bibfield
  {journal} {\bibinfo  {journal} {Phys. Rev. Lett.}\ }\textbf {\bibinfo
  {volume} {53}},\ \bibinfo {pages} {1515} (\bibinfo {year}
  {1984})}\BibitemShut {NoStop}%
\bibitem [{\citenamefont {Larkin}\ and\ \citenamefont
  {Ovchinnikov}(1969)}]{Larkin:1969}%
  \BibitemOpen
  \bibfield  {author} {\bibinfo {author} {\bibfnamefont {A.~I.}\ \bibnamefont
  {Larkin}}\ and\ \bibinfo {author} {\bibfnamefont {Y.~N.}\ \bibnamefont
  {Ovchinnikov}},\ }\bibfield  {title} {\enquote {\bibinfo {title}
  {{Quasiclassical method in the theory of superconductivity}},}\ }\href@noop
  {} {\bibfield  {journal} {\bibinfo  {journal} {JETP}\ }\textbf {\bibinfo
  {volume} {28}},\ \bibinfo {pages} {1200} (\bibinfo {year}
  {1969})}\BibitemShut {NoStop}%
\bibitem [{\citenamefont {Maldacena}\ \emph {et~al.}(2016)\citenamefont
  {Maldacena}, \citenamefont {Shenker},\ and\ \citenamefont
  {Stanford}}]{Maldacena:2016}%
  \BibitemOpen
  \bibfield  {author} {\bibinfo {author} {\bibfnamefont {J.}~\bibnamefont
  {Maldacena}}, \bibinfo {author} {\bibfnamefont {S.~H.}\ \bibnamefont
  {Shenker}}, \ and\ \bibinfo {author} {\bibfnamefont {D.}~\bibnamefont
  {Stanford}},\ }\bibfield  {title} {\enquote {\bibinfo {title} {A bound on
  chaos},}\ }\href {\doibase 10.1007/JHEP08(2016)106} {\bibfield  {journal}
  {\bibinfo  {journal} {J. High Energ. Phys.}\ }\textbf {\bibinfo {volume}
  {2016}},\ \bibinfo {pages} {106} (\bibinfo {year} {2016})}\BibitemShut
  {NoStop}%
\bibitem [{\citenamefont {Tsuji}\ \emph {et~al.}(2018)\citenamefont {Tsuji},
  \citenamefont {Shitara},\ and\ \citenamefont {Ueda}}]{Tsuji:2018}%
  \BibitemOpen
  \bibfield  {author} {\bibinfo {author} {\bibfnamefont {N.}~\bibnamefont
  {Tsuji}}, \bibinfo {author} {\bibfnamefont {T.}~\bibnamefont {Shitara}}, \
  and\ \bibinfo {author} {\bibfnamefont {M.}~\bibnamefont {Ueda}},\ }\bibfield
  {title} {\enquote {\bibinfo {title} {Bound on the exponential growth rate of
  out-of-time-ordered correlators},}\ }\href {\doibase
  10.1103/PhysRevE.98.012216} {\bibfield  {journal} {\bibinfo  {journal} {Phys.
  Rev. E}\ }\textbf {\bibinfo {volume} {98}},\ \bibinfo {pages} {012216}
  (\bibinfo {year} {2018})}\BibitemShut {NoStop}%
\bibitem [{\citenamefont {Yan}\ \emph {et~al.}(2020)\citenamefont {Yan},
  \citenamefont {Cincio},\ and\ \citenamefont {Zurek}}]{Yan:2020a}%
  \BibitemOpen
  \bibfield  {author} {\bibinfo {author} {\bibfnamefont {B.}~\bibnamefont
  {Yan}}, \bibinfo {author} {\bibfnamefont {L.}~\bibnamefont {Cincio}}, \ and\
  \bibinfo {author} {\bibfnamefont {W.~H.}\ \bibnamefont {Zurek}},\ }\bibfield
  {title} {\enquote {\bibinfo {title} {{Information Scrambling and Loschmidt
  Echo}},}\ }\href {\doibase 10.1103/PhysRevLett.124.160603} {\bibfield
  {journal} {\bibinfo  {journal} {Phys. Rev. Lett.}\ }\textbf {\bibinfo
  {volume} {124}},\ \bibinfo {pages} {160603} (\bibinfo {year}
  {2020})}\BibitemShut {NoStop}%
\bibitem [{\citenamefont {Yan}\ and\ \citenamefont
  {Sinitsyn}(2020)}]{Yan:2020}%
  \BibitemOpen
  \bibfield  {author} {\bibinfo {author} {\bibfnamefont {B.}~\bibnamefont
  {Yan}}\ and\ \bibinfo {author} {\bibfnamefont {N.~A.}\ \bibnamefont
  {Sinitsyn}},\ }\bibfield  {title} {\enquote {\bibinfo {title} {{Recovery of
  Damaged Information and the Out-of-Time-Ordered Correlators}},}\ }\href
  {\doibase 10.1103/PhysRevLett.125.040605} {\bibfield  {journal} {\bibinfo
  {journal} {Phys. Rev. Lett.}\ }\textbf {\bibinfo {volume} {125}},\ \bibinfo
  {pages} {040605} (\bibinfo {year} {2020})}\BibitemShut {NoStop}%
\bibitem [{\citenamefont {Landau}\ and\ \citenamefont
  {Lifshitz}(1980)}]{Landau:1980}%
  \BibitemOpen
  \bibfield  {author} {\bibinfo {author} {\bibfnamefont {L.D.}\ \bibnamefont
  {Landau}}\ and\ \bibinfo {author} {\bibfnamefont {E.M.}\ \bibnamefont
  {Lifshitz}},\ }\href@noop {} {\emph {\bibinfo {title} {Statistical Physics,
  Part I}}}\ (\bibinfo  {publisher} {Pergamon Press Ltd.},\ \bibinfo {address}
  {Oxford},\ \bibinfo {year} {1980})\BibitemShut {NoStop}%
\bibitem [{\citenamefont {Brink}\ \emph {et~al.}(1979)\citenamefont {Brink},
  \citenamefont {Neto},\ and\ \citenamefont {Weidenm{\"u}ller}}]{Brink:1979}%
  \BibitemOpen
  \bibfield  {author} {\bibinfo {author} {\bibfnamefont {D.M.}\ \bibnamefont
  {Brink}}, \bibinfo {author} {\bibfnamefont {J.}~\bibnamefont {Neto}}, \ and\
  \bibinfo {author} {\bibfnamefont {H.A.}\ \bibnamefont {Weidenm{\"u}ller}},\
  }\bibfield  {title} {\enquote {\bibinfo {title} {{Transport coefficients for
  deeply inelastic scattering from the Feynman path integral method}},}\ }\href
  {\doibase https://doi.org/10.1016/0370-2693(79)90190-4} {\bibfield  {journal}
  {\bibinfo  {journal} {Physics Letters B}\ }\textbf {\bibinfo {volume} {80}},\
  \bibinfo {pages} {170} (\bibinfo {year} {1979})}\BibitemShut {NoStop}%
\bibitem [{\citenamefont {Bulgac}\ \emph {et~al.}(1998)\citenamefont {Bulgac},
  \citenamefont {Do~Dang},\ and\ \citenamefont {Kusnezov}}]{Bulgac:1998}%
  \BibitemOpen
  \bibfield  {author} {\bibinfo {author} {\bibfnamefont {A.}~\bibnamefont
  {Bulgac}}, \bibinfo {author} {\bibfnamefont {G.}~\bibnamefont {Do~Dang}}, \
  and\ \bibinfo {author} {\bibfnamefont {D.}~\bibnamefont {Kusnezov}},\
  }\bibfield  {title} {\enquote {\bibinfo {title} {Dynamics of a simple quantum
  system in a complex environment},}\ }\href {\doibase 10.1103/PhysRevE.58.196}
  {\bibfield  {journal} {\bibinfo  {journal} {Phys. Rev. E}\ }\textbf {\bibinfo
  {volume} {58}},\ \bibinfo {pages} {196} (\bibinfo {year} {1998})}\BibitemShut
  {NoStop}%
\bibitem [{\citenamefont {Kusnezov}\ \emph {et~al.}(1999)\citenamefont
  {Kusnezov}, \citenamefont {Bulgac},\ and\ \citenamefont
  {Dang}}]{Kusnezov:1999}%
  \BibitemOpen
  \bibfield  {author} {\bibinfo {author} {\bibfnamefont {D.}~\bibnamefont
  {Kusnezov}}, \bibinfo {author} {\bibfnamefont {A.}~\bibnamefont {Bulgac}}, \
  and\ \bibinfo {author} {\bibfnamefont {G.~Do}\ \bibnamefont {Dang}},\
  }\bibfield  {title} {\enquote {\bibinfo {title} {{Quantum L\'evy Processes
  and Fractional Kinetics}},}\ }\href {\doibase 10.1103/PhysRevLett.82.1136}
  {\bibfield  {journal} {\bibinfo  {journal} {Phys. Rev. Lett.}\ }\textbf
  {\bibinfo {volume} {82}},\ \bibinfo {pages} {1136} (\bibinfo {year}
  {1999})}\BibitemShut {NoStop}%
\bibitem [{\citenamefont {Bethe}(1936)}]{Bethe:1936}%
  \BibitemOpen
  \bibfield  {author} {\bibinfo {author} {\bibfnamefont {H.~A.}\ \bibnamefont
  {Bethe}},\ }\bibfield  {title} {\enquote {\bibinfo {title} {An attempt to
  calculate the number of energy levels of a heavy nucleus},}\ }\href {\doibase
  10.1103/PhysRev.50.332} {\bibfield  {journal} {\bibinfo  {journal} {Phys.
  Rev.}\ }\textbf {\bibinfo {volume} {50}},\ \bibinfo {pages} {332} (\bibinfo
  {year} {1936})}\BibitemShut {NoStop}%
\bibitem [{\citenamefont {Deutsch}(1991)}]{Deutsch:1991}%
  \BibitemOpen
  \bibfield  {author} {\bibinfo {author} {\bibfnamefont {J.~M.}\ \bibnamefont
  {Deutsch}},\ }\bibfield  {title} {\enquote {\bibinfo {title} {Quantum
  statistical mechanics in a closed system},}\ }\href {\doibase
  10.1103/PhysRevA.43.2046} {\bibfield  {journal} {\bibinfo  {journal} {Phys.
  Rev. A}\ }\textbf {\bibinfo {volume} {43}},\ \bibinfo {pages} {2046}
  (\bibinfo {year} {1991})}\BibitemShut {NoStop}%
\bibitem [{\citenamefont {Srednicki}(1994)}]{Srednicki:1994}%
  \BibitemOpen
  \bibfield  {author} {\bibinfo {author} {\bibfnamefont {M.}~\bibnamefont
  {Srednicki}},\ }\bibfield  {title} {\enquote {\bibinfo {title} {Chaos and
  quantum thermalization},}\ }\href {\doibase 10.1103/PhysRevE.50.888}
  {\bibfield  {journal} {\bibinfo  {journal} {Phys. Rev. E}\ }\textbf {\bibinfo
  {volume} {50}},\ \bibinfo {pages} {888} (\bibinfo {year} {1994})}\BibitemShut
  {NoStop}%
\bibitem [{\citenamefont {Zelevinsky}\ \emph {et~al.}(1996)\citenamefont
  {Zelevinsky}, \citenamefont {Brown}, \citenamefont {Frazier},\ and\
  \citenamefont {Horoi}}]{Zelevinsky:1996}%
  \BibitemOpen
  \bibfield  {author} {\bibinfo {author} {\bibfnamefont {V.}~\bibnamefont
  {Zelevinsky}}, \bibinfo {author} {\bibfnamefont {B.~A.}\ \bibnamefont
  {Brown}}, \bibinfo {author} {\bibfnamefont {N.}~\bibnamefont {Frazier}}, \
  and\ \bibinfo {author} {\bibfnamefont {M.}~\bibnamefont {Horoi}},\ }\bibfield
   {title} {\enquote {\bibinfo {title} {The nuclear shell model as a testing
  ground for many-body chaos},}\ }\href {\doibase
  10.1016/S0370-1573(96)00007-5} {\bibfield  {journal} {\bibinfo  {journal}
  {Phys. Rep.}\ }\textbf {\bibinfo {volume} {276}},\ \bibinfo {pages} {85}
  (\bibinfo {year} {1996})}\BibitemShut {NoStop}%
\bibitem [{\citenamefont {Hohenberg}\ and\ \citenamefont
  {Kohn}(1964)}]{Hohenberg:1964}%
  \BibitemOpen
  \bibfield  {author} {\bibinfo {author} {\bibfnamefont {P.}~\bibnamefont
  {Hohenberg}}\ and\ \bibinfo {author} {\bibfnamefont {W.}~\bibnamefont
  {Kohn}},\ }\bibfield  {title} {\enquote {\bibinfo {title} {{Inhomogeneous
  Electron Gas}},}\ }\href {\doibase 10.1103/PhysRev.136.B864} {\bibfield
  {journal} {\bibinfo  {journal} {Phys. Rev.}\ }\textbf {\bibinfo {volume}
  {136}},\ \bibinfo {pages} {B864} (\bibinfo {year} {1964})}\BibitemShut
  {NoStop}%
\bibitem [{\citenamefont {Kohn}\ and\ \citenamefont
  {Sham}(1965)}]{Kohn:1965fk}%
  \BibitemOpen
  \bibfield  {author} {\bibinfo {author} {\bibfnamefont {W.}~\bibnamefont
  {Kohn}}\ and\ \bibinfo {author} {\bibfnamefont {L.~J.}\ \bibnamefont
  {Sham}},\ }\bibfield  {title} {\enquote {\bibinfo {title} {Self-consistent
  equations including exchange and correlation effects},}\ }\href {\doibase
  10.1103/PhysRev.140.A1133} {\bibfield  {journal} {\bibinfo  {journal} {Phys.
  Rev.}\ }\textbf {\bibinfo {volume} {140}},\ \bibinfo {pages} {A1133}
  (\bibinfo {year} {1965})}\BibitemShut {NoStop}%
\bibitem [{\citenamefont {Runge}\ and\ \citenamefont
  {Gross}(1984)}]{Runge:1984}%
  \BibitemOpen
  \bibfield  {author} {\bibinfo {author} {\bibfnamefont {E.}~\bibnamefont
  {Runge}}\ and\ \bibinfo {author} {\bibfnamefont {E.~K.~U.}\ \bibnamefont
  {Gross}},\ }\bibfield  {title} {\enquote {\bibinfo {title}
  {Density-functional theory for time-dependent systems},}\ }\href {\doibase
  10.1103/PhysRevLett.52.997} {\bibfield  {journal} {\bibinfo  {journal} {Phys.
  Rev. Lett.}\ }\textbf {\bibinfo {volume} {52}},\ \bibinfo {pages} {997--1000}
  (\bibinfo {year} {1984})}\BibitemShut {NoStop}%
\bibitem [{\citenamefont {Kohn}(1999)}]{Kohn:1999fk}%
  \BibitemOpen
  \bibfield  {author} {\bibinfo {author} {\bibfnamefont {W.}~\bibnamefont
  {Kohn}},\ }\bibfield  {title} {\enquote {\bibinfo {title} {{Nobel Lecture:
  Electronic structure of matter---wave functions and density functionals}},}\
  }\href {\doibase 10.1103/RevModPhys.71.1253} {\bibfield  {journal} {\bibinfo
  {journal} {Rev. Mod. Phys.}\ }\textbf {\bibinfo {volume} {71}},\ \bibinfo
  {pages} {1253} (\bibinfo {year} {1999})}\BibitemShut {NoStop}%
\bibitem [{\citenamefont {Dreizler}\ and\ \citenamefont
  {{Gross}}(1990)}]{Dreizler:1990lr}%
  \BibitemOpen
  \bibfield  {author} {\bibinfo {author} {\bibfnamefont {R.~M.}\ \bibnamefont
  {Dreizler}}\ and\ \bibinfo {author} {\bibfnamefont {E.~K.~U.}\ \bibnamefont
  {{Gross}}},\ }\href {\doibase 10.1007/978-3-642-86105-5} {\emph {\bibinfo
  {title} {{Density Functional Theory: An Approach to the Quantum Many--Body
  Problem}}}}\ (\bibinfo  {publisher} {Springer-Verlag},\ \bibinfo {address}
  {Berlin},\ \bibinfo {year} {1990})\BibitemShut {NoStop}%
\bibitem [{\citenamefont {Marques}\ \emph {et~al.}(2006)\citenamefont
  {Marques}, \citenamefont {Ullrich}, \citenamefont {Nogueira}, \citenamefont
  {Rubio}, \citenamefont {Burke},\ and\ \citenamefont {{Gross}}}]{Gross:2006}%
  \BibitemOpen
  \bibinfo {editor} {\bibfnamefont {M.~A.~L.}\ \bibnamefont {Marques}},
  \bibinfo {editor} {\bibfnamefont {C.~A.}\ \bibnamefont {Ullrich}}, \bibinfo
  {editor} {\bibfnamefont {F.}~\bibnamefont {Nogueira}}, \bibinfo {editor}
  {\bibfnamefont {A.}~\bibnamefont {Rubio}}, \bibinfo {editor} {\bibfnamefont
  {K.}~\bibnamefont {Burke}}, \ and\ \bibinfo {editor} {\bibfnamefont
  {E.~K.~U.}\ \bibnamefont {{Gross}}},\ eds.,\ \href {\doibase
  10.1007/b11767107} {\emph {\bibinfo {title} {Time-Dependent Density
  Functional Theory}}},\ \bibinfo {series} {Lecture Notes in Physics}, Vol.\
  \bibinfo {volume} {706}\ (\bibinfo  {publisher} {Springer-Verlag},\ \bibinfo
  {address} {Berlin},\ \bibinfo {year} {2006})\BibitemShut {NoStop}%
\bibitem [{\citenamefont {Marques}\ \emph {et~al.}(2012)\citenamefont
  {Marques}, \citenamefont {Maitra}, \citenamefont {Nogueira}, \citenamefont
  {{Gross}},\ and\ \citenamefont {Rubio}}]{Gross:2012}%
  \BibitemOpen
  \bibinfo {editor} {\bibfnamefont {M.~A.~L.}\ \bibnamefont {Marques}},
  \bibinfo {editor} {\bibfnamefont {N.~T.}\ \bibnamefont {Maitra}}, \bibinfo
  {editor} {\bibfnamefont {F.~M.~S.}\ \bibnamefont {Nogueira}}, \bibinfo
  {editor} {\bibfnamefont {E.~K.~U.}\ \bibnamefont {{Gross}}}, \ and\ \bibinfo
  {editor} {\bibfnamefont {A.}~\bibnamefont {Rubio}},\ eds.,\ \href {\doibase
  10.1007/978-3-642-23518-4} {\emph {\bibinfo {title} {Fundamentals of
  Time-Dependent Density Functional Theory}}},\ \bibinfo {series} {Lecture
  Notes in Physics}, Vol.\ \bibinfo {volume} {837}\ (\bibinfo  {publisher}
  {Springer},\ \bibinfo {address} {Heidelberg},\ \bibinfo {year}
  {2012})\BibitemShut {NoStop}%
\bibitem [{\citenamefont {Balian}\ and\ \citenamefont
  {V\'en\'eroni}(1989)}]{Balian:1989}%
  \BibitemOpen
  \bibfield  {author} {\bibinfo {author} {\bibfnamefont {R.}~\bibnamefont
  {Balian}}\ and\ \bibinfo {author} {\bibfnamefont {M.}~\bibnamefont
  {V\'en\'eroni}},\ }\bibfield  {title} {\enquote {\bibinfo {title} {{Lyapunov
  stability and Poison structure of the thermal TDHF and RPA}},}\ }\href
  {\doibase 10.1016/0003-4916(89)90247-9} {\bibfield  {journal} {\bibinfo
  {journal} {Ann. Phys.}\ }\textbf {\bibinfo {volume} {195}},\ \bibinfo {pages}
  {324} (\bibinfo {year} {1989})}\BibitemShut {NoStop}%
\bibitem [{\citenamefont {Bulgac}\ \emph
  {et~al.}(2016{\natexlab{a}})\citenamefont {Bulgac}, \citenamefont {Forbes},\
  and\ \citenamefont {Wlaz{\l}owski}}]{Bulgac:2016x}%
  \BibitemOpen
  \bibfield  {author} {\bibinfo {author} {\bibfnamefont {A.}~\bibnamefont
  {Bulgac}}, \bibinfo {author} {\bibfnamefont {M.~M.}\ \bibnamefont {Forbes}},
  \ and\ \bibinfo {author} {\bibfnamefont {W.}~\bibnamefont {Wlaz{\l}owski}},\
  }\bibfield  {title} {\enquote {\bibinfo {title} {Towards quantum turbulence
  in cold atomic fermionic superfluids},}\ }\href {\doibase
  10.1088/1361-6455/50/1/014001} {\bibfield  {journal} {\bibinfo  {journal} {J.
  Phys. B: At. Mol. Opt. Phys.}\ }\textbf {\bibinfo {volume} {50}},\ \bibinfo
  {pages} {014001} (\bibinfo {year} {2016}{\natexlab{a}})}\BibitemShut
  {NoStop}%
\bibitem [{\citenamefont {Bulgac}(2019)}]{Bulgac:2019}%
  \BibitemOpen
  \bibfield  {author} {\bibinfo {author} {\bibfnamefont {A.}~\bibnamefont
  {Bulgac}},\ }\bibfield  {title} {\enquote {\bibinfo {title} {{Time-Dependent
  Density Functional Theory for Fermionic Superfluids: from Cold Atomic gases,
  to Nuclei and Neutron Star Crust}},}\ }\href {\doibase
  10.1002/pssb.201800592} {\bibfield  {journal} {\bibinfo  {journal} {Physica
  Status Solidi B}\ }\textbf {\bibinfo {volume} {256}},\ \bibinfo {pages}
  {1800592} (\bibinfo {year} {2019})}\BibitemShut {NoStop}%
\bibitem [{\citenamefont {Jin}\ \emph {et~al.}(2021)\citenamefont {Jin},
  \citenamefont {Roche}, \citenamefont {Stetcu}, \citenamefont {Abdurrahman},\
  and\ \citenamefont {Bulgac}}]{Shi:2020}%
  \BibitemOpen
  \bibfield  {author} {\bibinfo {author} {\bibfnamefont {S.}~\bibnamefont
  {Jin}}, \bibinfo {author} {\bibfnamefont {K.~J.}\ \bibnamefont {Roche}},
  \bibinfo {author} {\bibfnamefont {I.}~\bibnamefont {Stetcu}}, \bibinfo
  {author} {\bibfnamefont {I.}~\bibnamefont {Abdurrahman}}, \ and\ \bibinfo
  {author} {\bibfnamefont {A.}~\bibnamefont {Bulgac}},\ }\bibfield  {title}
  {\enquote {\bibinfo {title} {{The LISE package: solvers for static and
  time-dependent superfluid local density approximation equations in three
  dimensions}},}\ }\href {\doibase 10.1016/j.cpc.2021.108130} {\bibfield
  {journal} {\bibinfo  {journal} {Comput. Phys. Commun.}\ }\textbf {\bibinfo
  {volume} {269}},\ \bibinfo {pages} {108130} (\bibinfo {year}
  {2021})}\BibitemShut {NoStop}%
\bibitem [{\citenamefont {Ring}\ and\ \citenamefont
  {Schuck}(2004)}]{Ring:2004}%
  \BibitemOpen
  \bibfield  {author} {\bibinfo {author} {\bibfnamefont {P.}~\bibnamefont
  {Ring}}\ and\ \bibinfo {author} {\bibfnamefont {P.}~\bibnamefont {Schuck}},\
  }\href@noop {} {\emph {\bibinfo {title} {{The Nuclear Many-Body Problem}}}},\
  \bibinfo {edition} {1st}\ ed.,\ \bibinfo {series} {Theoretical and
  Mathematical Physics Series}\ No.~\bibinfo {number} {17}\ (\bibinfo
  {publisher} {Springer-Verlag},\ \bibinfo {address} {Berlin Heidelberg New
  York},\ \bibinfo {year} {2004})\BibitemShut {NoStop}%
\bibitem [{\citenamefont {Bertsch}()}]{Bertsch:1999}%
  \BibitemOpen
  \bibfield  {author} {\bibinfo {author} {\bibfnamefont {G.~F.}\ \bibnamefont
  {Bertsch}},\ }\href@noop {} {\enquote {\bibinfo {title} {The {M}any-{B}ody
  {C}hallenge {P}roblem {(\textsc{mbx})} formulated in 1999},}\ }\bibinfo
  {howpublished} {see also Ref.~\cite{Baker:1999}.}\BibitemShut {Stop}%
\bibitem [{\citenamefont {Baker}(1999)}]{Baker:1999}%
  \BibitemOpen
  \bibfield  {author} {\bibinfo {author} {\bibfnamefont {G.~A.}\ \bibnamefont
  {Baker}, \bibfnamefont {Jr.}},\ }\bibfield  {title} {\enquote {\bibinfo
  {title} {Neutron matter model},}\ }\href {\doibase
  10.1103/PhysRevC.60.054311} {\bibfield  {journal} {\bibinfo  {journal} {Phys.
  Rev. C}\ }\textbf {\bibinfo {volume} {60}},\ \bibinfo {pages} {054311}
  (\bibinfo {year} {1999})}\BibitemShut {NoStop}%
\bibitem [{\citenamefont {Zwerger}(2012)}]{Zwerger:2011}%
  \BibitemOpen
  \bibinfo {editor} {\bibfnamefont {W.}~\bibnamefont {Zwerger}},\ ed.,\ \href
  {\doibase 10.1007/978-3-642-21978-8} {\emph {\bibinfo {title} {The BCS--BEC
  Crossover and the Unitary {Fermi} Gas}}},\ \bibinfo {series} {Lecture Notes
  in Physics}, Vol.\ \bibinfo {volume} {836}\ (\bibinfo  {publisher}
  {Springer-Verlag},\ \bibinfo {address} {Berlin Heidelberg},\ \bibinfo {year}
  {2012})\BibitemShut {NoStop}%
\bibitem [{\citenamefont {Carlson}\ \emph {et~al.}(2011)\citenamefont
  {Carlson}, \citenamefont {Gandolfi}, \citenamefont {Schmidt},\ and\
  \citenamefont {Zhang}}]{Carlson:2011}%
  \BibitemOpen
  \bibfield  {author} {\bibinfo {author} {\bibfnamefont {J.}~\bibnamefont
  {Carlson}}, \bibinfo {author} {\bibfnamefont {S.}~\bibnamefont {Gandolfi}},
  \bibinfo {author} {\bibfnamefont {K.~E.}\ \bibnamefont {Schmidt}}, \ and\
  \bibinfo {author} {\bibfnamefont {S.}~\bibnamefont {Zhang}},\ }\bibfield
  {title} {\enquote {\bibinfo {title} {Auxiliary field quantum {Monte} {Carlo}
  for strongly paired fermions},}\ }\href {\doibase 10.1103/PhysRevA.84.061602}
  {\bibfield  {journal} {\bibinfo  {journal} {Phys. Rev. A}\ }\textbf {\bibinfo
  {volume} {84}},\ \bibinfo {pages} {061602} (\bibinfo {year}
  {2011})}\BibitemShut {NoStop}%
\bibitem [{\citenamefont {Ku}\ \emph {et~al.}(2012)\citenamefont {Ku},
  \citenamefont {Sommer}, \citenamefont {Cheuk},\ and\ \citenamefont
  {Zwierlein}}]{Ku:2011}%
  \BibitemOpen
  \bibfield  {author} {\bibinfo {author} {\bibfnamefont {M.~J.~H.}\
  \bibnamefont {Ku}}, \bibinfo {author} {\bibfnamefont {A.~T.}\ \bibnamefont
  {Sommer}}, \bibinfo {author} {\bibfnamefont {L.~W.}\ \bibnamefont {Cheuk}}, \
  and\ \bibinfo {author} {\bibfnamefont {M.~W.}\ \bibnamefont {Zwierlein}},\
  }\bibfield  {title} {\enquote {\bibinfo {title} {{R}evealing the {S}uperfluid
  {L}ambda {T}ransition in the {U}niversal {T}hermodynamics of a {U}nitary
  {Fermi} {G}as},}\ }\href {\doibase 10.1126/science.1214987} {\bibfield
  {journal} {\bibinfo  {journal} {Science}\ }\textbf {\bibinfo {volume}
  {335}},\ \bibinfo {pages} {563} (\bibinfo {year} {2012})}\BibitemShut
  {NoStop}%
\bibitem [{\citenamefont {Bulgac}(2007)}]{Bulgac:2007}%
  \BibitemOpen
  \bibfield  {author} {\bibinfo {author} {\bibfnamefont {A.}~\bibnamefont
  {Bulgac}},\ }\bibfield  {title} {\enquote {\bibinfo {title}
  {{Local-density-functional theory for superfluid fermionic systems: The
  unitary gas}},}\ }\href {\doibase 10.1103/PhysRevA.76.040502} {\bibfield
  {journal} {\bibinfo  {journal} {Phys. Rev. A}\ }\textbf {\bibinfo {volume}
  {76}},\ \bibinfo {pages} {040502} (\bibinfo {year} {2007})}\BibitemShut
  {NoStop}%
\bibitem [{\citenamefont {Bulgac}\ \emph {et~al.}(2012)\citenamefont {Bulgac},
  \citenamefont {Forbes},\ and\ \citenamefont {Magierski}}]{Bulgac:2011a}%
  \BibitemOpen
  \bibfield  {author} {\bibinfo {author} {\bibfnamefont {A.}~\bibnamefont
  {Bulgac}}, \bibinfo {author} {\bibfnamefont {M.~M.}\ \bibnamefont {Forbes}},
  \ and\ \bibinfo {author} {\bibfnamefont {P.}~\bibnamefont {Magierski}},\
  }\enquote {\bibinfo {title} {The unitary {Fermi} gas: From {Monte} {Carlo} to
  density functionals},}\ Chap.~\bibinfo {chapter} {9}, pp.\ \bibinfo {pages}
  {127 -- 191},\ vol.\ \bibinfo {volume} {836}\ of\  \cite{Zwerger:2011}
  (\bibinfo {year} {2012})\BibitemShut {NoStop}%
\bibitem [{\citenamefont {Yuzbashyan}\ and\ \citenamefont
  {Dzero}(2006)}]{Yuzbashyan:2006}%
  \BibitemOpen
  \bibfield  {author} {\bibinfo {author} {\bibfnamefont {E.~A.}\ \bibnamefont
  {Yuzbashyan}}\ and\ \bibinfo {author} {\bibfnamefont {M.}~\bibnamefont
  {Dzero}},\ }\bibfield  {title} {\enquote {\bibinfo {title} {Dynamical
  vanishing of the order parameter in a fermionic condensate},}\ }\href
  {\doibase 10.1103/PhysRevLett.96.230404} {\bibfield  {journal} {\bibinfo
  {journal} {Phys. Rev. Lett.}\ }\textbf {\bibinfo {volume} {96}},\ \bibinfo
  {pages} {230404} (\bibinfo {year} {2006})}\BibitemShut {NoStop}%
\bibitem [{\citenamefont {Yuzbashyan}\ \emph {et~al.}(2006)\citenamefont
  {Yuzbashyan}, \citenamefont {Tsyplyatyev},\ and\ \citenamefont
  {Altshuler}}]{Yuzbashyan:2006a}%
  \BibitemOpen
  \bibfield  {author} {\bibinfo {author} {\bibfnamefont {E.~A.}\ \bibnamefont
  {Yuzbashyan}}, \bibinfo {author} {\bibfnamefont {O.}~\bibnamefont
  {Tsyplyatyev}}, \ and\ \bibinfo {author} {\bibfnamefont {B.~L.}\ \bibnamefont
  {Altshuler}},\ }\bibfield  {title} {\enquote {\bibinfo {title} {Relaxation
  and persistent oscillations of the order parameter in fermionic
  condensates},}\ }\href {\doibase 10.1103/PhysRevLett.96.097005} {\bibfield
  {journal} {\bibinfo  {journal} {Phys. Rev. Lett.}\ }\textbf {\bibinfo
  {volume} {96}},\ \bibinfo {pages} {097005} (\bibinfo {year}
  {2006})}\BibitemShut {NoStop}%
\bibitem [{\citenamefont {Yuzbashyan}(2008)}]{Yuzbashyan:2008}%
  \BibitemOpen
  \bibfield  {author} {\bibinfo {author} {\bibfnamefont {E.~A.}\ \bibnamefont
  {Yuzbashyan}},\ }\bibfield  {title} {\enquote {\bibinfo {title} {Normal and
  anomalous solitons in the theory of dynamical cooper pairing},}\ }\href
  {\doibase 10.1103/PhysRevB.78.184507} {\bibfield  {journal} {\bibinfo
  {journal} {Phys. Rev. B}\ }\textbf {\bibinfo {volume} {78}},\ \bibinfo
  {pages} {184507} (\bibinfo {year} {2008})}\BibitemShut {NoStop}%
\bibitem [{\citenamefont {Bulgac}\ and\ \citenamefont
  {Yoon}(2009)}]{Bulgac:2009}%
  \BibitemOpen
  \bibfield  {author} {\bibinfo {author} {\bibfnamefont {A.}~\bibnamefont
  {Bulgac}}\ and\ \bibinfo {author} {\bibfnamefont {S.}~\bibnamefont {Yoon}},\
  }\bibfield  {title} {\enquote {\bibinfo {title} {{Large Amplitude Dynamics of
  the Pairing Correlations in a Unitary Fermi Gas}},}\ }\href {\doibase
  10.1103/PhysRevLett.102.085302} {\bibfield  {journal} {\bibinfo  {journal}
  {Phys. Rev. Lett.}\ }\textbf {\bibinfo {volume} {102}},\ \bibinfo {pages}
  {085302} (\bibinfo {year} {2009})}\BibitemShut {NoStop}%
\bibitem [{\citenamefont {Bulgac}\ \emph {et~al.}(2011)\citenamefont {Bulgac},
  \citenamefont {Luo}, \citenamefont {Magierski}, \citenamefont {Roche},\ and\
  \citenamefont {Yu}}]{Bulgac:2011}%
  \BibitemOpen
  \bibfield  {author} {\bibinfo {author} {\bibfnamefont {A.}~\bibnamefont
  {Bulgac}}, \bibinfo {author} {\bibfnamefont {Y.-L.}\ \bibnamefont {Luo}},
  \bibinfo {author} {\bibfnamefont {P.}~\bibnamefont {Magierski}}, \bibinfo
  {author} {\bibfnamefont {K.~J.}\ \bibnamefont {Roche}}, \ and\ \bibinfo
  {author} {\bibfnamefont {Y.}~\bibnamefont {Yu}},\ }\bibfield  {title}
  {\enquote {\bibinfo {title} {{Real-Time Dynamics of Quantized Vortices in a
  Unitary Fermi Superfluid}},}\ }\href {\doibase 10.1126/science.1201968}
  {\bibfield  {journal} {\bibinfo  {journal} {Science}\ }\textbf {\bibinfo
  {volume} {332}},\ \bibinfo {pages} {1288} (\bibinfo {year}
  {2011})}\BibitemShut {NoStop}%
\bibitem [{\citenamefont {Bulgac}\ \emph {et~al.}(2014)\citenamefont {Bulgac},
  \citenamefont {Forbes}, \citenamefont {Kelley}, \citenamefont {Roche},\ and\
  \citenamefont {Wlaz\l{}owski}}]{Bulgac:2014}%
  \BibitemOpen
  \bibfield  {author} {\bibinfo {author} {\bibfnamefont {A.}~\bibnamefont
  {Bulgac}}, \bibinfo {author} {\bibfnamefont {M.~M.}\ \bibnamefont {Forbes}},
  \bibinfo {author} {\bibfnamefont {M.~M.}\ \bibnamefont {Kelley}}, \bibinfo
  {author} {\bibfnamefont {K.~J.}\ \bibnamefont {Roche}}, \ and\ \bibinfo
  {author} {\bibfnamefont {G.}~\bibnamefont {Wlaz\l{}owski}},\ }\bibfield
  {title} {\enquote {\bibinfo {title} {Quantized superfluid vortex rings in the
  unitary fermi gas},}\ }\href {\doibase 10.1103/PhysRevLett.112.025301}
  {\bibfield  {journal} {\bibinfo  {journal} {Phys. Rev. Lett.}\ }\textbf
  {\bibinfo {volume} {112}},\ \bibinfo {pages} {025301} (\bibinfo {year}
  {2014})}\BibitemShut {NoStop}%
\bibitem [{\citenamefont {Wlaz\l{}owski}\ \emph
  {et~al.}(2015{\natexlab{a}})\citenamefont {Wlaz\l{}owski}, \citenamefont
  {Bulgac}, \citenamefont {Forbes},\ and\ \citenamefont
  {Roche}}]{Wlazlowski:2015}%
  \BibitemOpen
  \bibfield  {author} {\bibinfo {author} {\bibfnamefont {G.}~\bibnamefont
  {Wlaz\l{}owski}}, \bibinfo {author} {\bibfnamefont {A.}~\bibnamefont
  {Bulgac}}, \bibinfo {author} {\bibfnamefont {M.~M.}\ \bibnamefont {Forbes}},
  \ and\ \bibinfo {author} {\bibfnamefont {K.~J.}\ \bibnamefont {Roche}},\
  }\bibfield  {title} {\enquote {\bibinfo {title} {{Life cycle of superfluid
  vortices and quantum turbulence in the unitary Fermi gas}},}\ }\href
  {\doibase 10.1103/PhysRevA.91.031602} {\bibfield  {journal} {\bibinfo
  {journal} {Phys. Rev. A}\ }\textbf {\bibinfo {volume} {91}},\ \bibinfo
  {pages} {031602} (\bibinfo {year} {2015}{\natexlab{a}})}\BibitemShut
  {NoStop}%
\bibitem [{\citenamefont {Wlaz\l{}owski}\ \emph {et~al.}(2018)\citenamefont
  {Wlaz\l{}owski}, \citenamefont {Sekizawa}, \citenamefont {Marchwiany},\ and\
  \citenamefont {Magierski}}]{Wlazlowski:2018}%
  \BibitemOpen
  \bibfield  {author} {\bibinfo {author} {\bibfnamefont {G.}~\bibnamefont
  {Wlaz\l{}owski}}, \bibinfo {author} {\bibfnamefont {K.}~\bibnamefont
  {Sekizawa}}, \bibinfo {author} {\bibfnamefont {M.}~\bibnamefont
  {Marchwiany}}, \ and\ \bibinfo {author} {\bibfnamefont {P.}~\bibnamefont
  {Magierski}},\ }\bibfield  {title} {\enquote {\bibinfo {title} {{Suppressed
  Solitonic Cascade in Spin-Imbalanced Superfluid Fermi Gas}},}\ }\href
  {\doibase 10.1103/PhysRevLett.120.253002} {\bibfield  {journal} {\bibinfo
  {journal} {Phys. Rev. Lett.}\ }\textbf {\bibinfo {volume} {120}},\ \bibinfo
  {pages} {253002} (\bibinfo {year} {2018})}\BibitemShut {NoStop}%
\bibitem [{\citenamefont {Yefsah}\ \emph {et~al.}(2013)\citenamefont {Yefsah},
  \citenamefont {Sommer}, \citenamefont {Ku}, \citenamefont {Cheuk},
  \citenamefont {Ji}, \citenamefont {Bakr},\ and\ \citenamefont
  {Zwierlein}}]{Yefsah:2013}%
  \BibitemOpen
  \bibfield  {author} {\bibinfo {author} {\bibfnamefont {T.}~\bibnamefont
  {Yefsah}}, \bibinfo {author} {\bibfnamefont {A.~T.}\ \bibnamefont {Sommer}},
  \bibinfo {author} {\bibfnamefont {M.~J.~H.}\ \bibnamefont {Ku}}, \bibinfo
  {author} {\bibfnamefont {L.~W.}\ \bibnamefont {Cheuk}}, \bibinfo {author}
  {\bibfnamefont {W.}~\bibnamefont {Ji}}, \bibinfo {author} {\bibfnamefont
  {W.~S.}\ \bibnamefont {Bakr}}, \ and\ \bibinfo {author} {\bibfnamefont
  {M.~W.}\ \bibnamefont {Zwierlein}},\ }\bibfield  {title} {\enquote {\bibinfo
  {title} {{Heavy Solitons in a Fermionic Superfluid}},}\ }\href {\doibase
  10.1038/nature12338} {\bibfield  {journal} {\bibinfo  {journal} {Nature}\
  }\textbf {\bibinfo {volume} {499}},\ \bibinfo {pages} {426--430} (\bibinfo
  {year} {2013})}\BibitemShut {NoStop}%
\bibitem [{\citenamefont {Ku}\ \emph {et~al.}(2014)\citenamefont {Ku},
  \citenamefont {Ji}, \citenamefont {Mukherjee}, \citenamefont
  {Guardado-Sanchez}, \citenamefont {Cheuk}, \citenamefont {Yefsah},\ and\
  \citenamefont {Zwierlein}}]{Ku:2014}%
  \BibitemOpen
  \bibfield  {author} {\bibinfo {author} {\bibfnamefont {M.~J.~H.}\
  \bibnamefont {Ku}}, \bibinfo {author} {\bibfnamefont {W.}~\bibnamefont {Ji}},
  \bibinfo {author} {\bibfnamefont {B.}~\bibnamefont {Mukherjee}}, \bibinfo
  {author} {\bibfnamefont {E.}~\bibnamefont {Guardado-Sanchez}}, \bibinfo
  {author} {\bibfnamefont {L.~W.}\ \bibnamefont {Cheuk}}, \bibinfo {author}
  {\bibfnamefont {T.}~\bibnamefont {Yefsah}}, \ and\ \bibinfo {author}
  {\bibfnamefont {M.~W.}\ \bibnamefont {Zwierlein}},\ }\bibfield  {title}
  {\enquote {\bibinfo {title} {{Motion of a Solitonic Vortex in the {BEC-BCS}
  Crossover}},}\ }\href {\doibase 10.1103/PhysRevLett.113.065301} {\bibfield
  {journal} {\bibinfo  {journal} {Phys. Rev. Lett.}\ }\textbf {\bibinfo
  {volume} {113}},\ \bibinfo {pages} {065301} (\bibinfo {year}
  {2014})}\BibitemShut {NoStop}%
\bibitem [{\citenamefont {Stetcu}\ \emph {et~al.}(2011)\citenamefont {Stetcu},
  \citenamefont {Bulgac}, \citenamefont {Magierski},\ and\ \citenamefont
  {Roche}}]{Stetcu:2011}%
  \BibitemOpen
  \bibfield  {author} {\bibinfo {author} {\bibfnamefont {I.}~\bibnamefont
  {Stetcu}}, \bibinfo {author} {\bibfnamefont {A.}~\bibnamefont {Bulgac}},
  \bibinfo {author} {\bibfnamefont {P.}~\bibnamefont {Magierski}}, \ and\
  \bibinfo {author} {\bibfnamefont {K.~J.}\ \bibnamefont {Roche}},\ }\bibfield
  {title} {\enquote {\bibinfo {title} {Isovector giant dipole resonance from
  the 3d time-dependent density functional theory for superfluid nuclei},}\
  }\href {\doibase 10.1103/PhysRevC.84.051309} {\bibfield  {journal} {\bibinfo
  {journal} {Phys. Rev. C}\ }\textbf {\bibinfo {volume} {84}},\ \bibinfo
  {pages} {051309} (\bibinfo {year} {2011})}\BibitemShut {NoStop}%
\bibitem [{\citenamefont {Stetcu}\ \emph {et~al.}(2015)\citenamefont {Stetcu},
  \citenamefont {Bertulani}, \citenamefont {Bulgac}, \citenamefont
  {Magierski},\ and\ \citenamefont {Roche}}]{Stetcu:2014}%
  \BibitemOpen
  \bibfield  {author} {\bibinfo {author} {\bibfnamefont {I.}~\bibnamefont
  {Stetcu}}, \bibinfo {author} {\bibfnamefont {C.}~\bibnamefont {Bertulani}},
  \bibinfo {author} {\bibfnamefont {A.}~\bibnamefont {Bulgac}}, \bibinfo
  {author} {\bibfnamefont {P.}~\bibnamefont {Magierski}}, \ and\ \bibinfo
  {author} {\bibfnamefont {K.~J.}\ \bibnamefont {Roche}},\ }\bibfield  {title}
  {\enquote {\bibinfo {title} {{R}elativistic {C}oulomb excitation within
  {T}ime {D}ependent {S}uperfluid {L}ocal {D}ensity {A}pproximation},}\ }\href
  {\doibase 10.1103/PhysRevLett.114.012701} {\bibfield  {journal} {\bibinfo
  {journal} {Phys. Rev. Lett.}\ }\textbf {\bibinfo {volume} {114}},\ \bibinfo
  {pages} {012701} (\bibinfo {year} {2015})}\BibitemShut {NoStop}%
\bibitem [{\citenamefont {Bulgac}\ \emph
  {et~al.}(2016{\natexlab{b}})\citenamefont {Bulgac}, \citenamefont
  {Magierski}, \citenamefont {Roche},\ and\ \citenamefont
  {Stetcu}}]{Bulgac:2016}%
  \BibitemOpen
  \bibfield  {author} {\bibinfo {author} {\bibfnamefont {A.}~\bibnamefont
  {Bulgac}}, \bibinfo {author} {\bibfnamefont {P.}~\bibnamefont {Magierski}},
  \bibinfo {author} {\bibfnamefont {K.~J.}\ \bibnamefont {Roche}}, \ and\
  \bibinfo {author} {\bibfnamefont {I.}~\bibnamefont {Stetcu}},\ }\bibfield
  {title} {\enquote {\bibinfo {title} {{Induced Fission of $^{240}\mathrm{Pu}$
  within a Real-Time Microscopic Framework}},}\ }\href {\doibase
  10.1103/PhysRevLett.116.122504} {\bibfield  {journal} {\bibinfo  {journal}
  {Phys. Rev. Lett.}\ }\textbf {\bibinfo {volume} {116}},\ \bibinfo {pages}
  {122504} (\bibinfo {year} {2016}{\natexlab{b}})}\BibitemShut {NoStop}%
\bibitem [{\citenamefont {Wlaz\l{}owski}\ \emph {et~al.}(2016)\citenamefont
  {Wlaz\l{}owski}, \citenamefont {Sekizawa}, \citenamefont {Magierski},
  \citenamefont {Bulgac},\ and\ \citenamefont {Forbes}}]{Wlazlowski:2016}%
  \BibitemOpen
  \bibfield  {author} {\bibinfo {author} {\bibfnamefont {G.}~\bibnamefont
  {Wlaz\l{}owski}}, \bibinfo {author} {\bibfnamefont {K.}~\bibnamefont
  {Sekizawa}}, \bibinfo {author} {\bibfnamefont {P.}~\bibnamefont {Magierski}},
  \bibinfo {author} {\bibfnamefont {A.}~\bibnamefont {Bulgac}}, \ and\ \bibinfo
  {author} {\bibfnamefont {M.~M.}\ \bibnamefont {Forbes}},\ }\bibfield  {title}
  {\enquote {\bibinfo {title} {{Vortex Pinning and Dynamics in the Neutron Star
  Crust}},}\ }\href {\doibase 10.1103/PhysRevLett.117.232701} {\bibfield
  {journal} {\bibinfo  {journal} {Phys. Rev. Lett.}\ }\textbf {\bibinfo
  {volume} {117}},\ \bibinfo {pages} {232701} (\bibinfo {year}
  {2016})}\BibitemShut {NoStop}%
\bibitem [{\citenamefont {Bulgac}\ and\ \citenamefont
  {Jin}(2017)}]{Bulgac:2017}%
  \BibitemOpen
  \bibfield  {author} {\bibinfo {author} {\bibfnamefont {A.}~\bibnamefont
  {Bulgac}}\ and\ \bibinfo {author} {\bibfnamefont {S.}~\bibnamefont {Jin}},\
  }\bibfield  {title} {\enquote {\bibinfo {title} {{Dynamics of Fragmented
  Condensates and Macroscopic Entanglement}},}\ }\href {\doibase
  10.1103/PhysRevLett.119.052501} {\bibfield  {journal} {\bibinfo  {journal}
  {Phys. Rev. Lett.}\ }\textbf {\bibinfo {volume} {119}},\ \bibinfo {pages}
  {052501} (\bibinfo {year} {2017})}\BibitemShut {NoStop}%
\bibitem [{\citenamefont {Bulgac}\ \emph {et~al.}(2019)\citenamefont {Bulgac},
  \citenamefont {Jin}, \citenamefont {Roche}, \citenamefont {Schunck},\ and\
  \citenamefont {Stetcu}}]{Bulgac:2019b}%
  \BibitemOpen
  \bibfield  {author} {\bibinfo {author} {\bibfnamefont {A.}~\bibnamefont
  {Bulgac}}, \bibinfo {author} {\bibfnamefont {S.}~\bibnamefont {Jin}},
  \bibinfo {author} {\bibfnamefont {K.~J.}\ \bibnamefont {Roche}}, \bibinfo
  {author} {\bibfnamefont {N.}~\bibnamefont {Schunck}}, \ and\ \bibinfo
  {author} {\bibfnamefont {I.}~\bibnamefont {Stetcu}},\ }\bibfield  {title}
  {\enquote {\bibinfo {title} {Fission dynamics of $^{240}\mathrm{Pu}$ from
  saddle to scission and beyond},}\ }\href {\doibase
  10.1103/PhysRevC.100.034615} {\bibfield  {journal} {\bibinfo  {journal}
  {Phys. Rev. C}\ }\textbf {\bibinfo {volume} {100}},\ \bibinfo {pages}
  {034615} (\bibinfo {year} {2019})}\BibitemShut {NoStop}%
\bibitem [{\citenamefont {Bulgac}\ \emph {et~al.}(2020)\citenamefont {Bulgac},
  \citenamefont {Jin},\ and\ \citenamefont {Stetcu}}]{Bulgac:2020}%
  \BibitemOpen
  \bibfield  {author} {\bibinfo {author} {\bibfnamefont {A.}~\bibnamefont
  {Bulgac}}, \bibinfo {author} {\bibfnamefont {S.}~\bibnamefont {Jin}}, \ and\
  \bibinfo {author} {\bibfnamefont {I.}~\bibnamefont {Stetcu}},\ }\bibfield
  {title} {\enquote {\bibinfo {title} {{Nuclear Fission Dynamics: Past,
  Present, Needs, and Future}},}\ }\href {\doibase 10.3389/fphy.2020.00063}
  {\bibfield  {journal} {\bibinfo  {journal} {{Frontiers in Physics}}\ }\textbf
  {\bibinfo {volume} {8}},\ \bibinfo {pages} {63} (\bibinfo {year}
  {2020})}\BibitemShut {NoStop}%
\bibitem [{\citenamefont {Feynman}(1955)}]{Feynman:1955}%
  \BibitemOpen
  \bibfield  {author} {\bibinfo {author} {\bibfnamefont {R.P.}\ \bibnamefont
  {Feynman}},\ }\bibfield  {title} {\enquote {\bibinfo {title} {Application of
  quantum mechanics to liquid helium},}\ }\href {\doibase
  10.1016/S0079-6417(08)60077-3} {\bibfield  {journal} {\bibinfo  {journal}
  {Prog. Low Temp. Phys.}\ }\textbf {\bibinfo {volume} {1}},\ \bibinfo {pages}
  {17} (\bibinfo {year} {1955})}\BibitemShut {NoStop}%
\bibitem [{\citenamefont {Barenghi}\ \emph {et~al.}(2001)\citenamefont
  {Barenghi}, \citenamefont {Donnelly},\ and\ \citenamefont {Vinen}}]{VFM2001}%
  \BibitemOpen
  \bibinfo {editor} {\bibfnamefont {C.~F.}\ \bibnamefont {Barenghi}}, \bibinfo
  {editor} {\bibfnamefont {R.~J.}\ \bibnamefont {Donnelly}}, \ and\ \bibinfo
  {editor} {\bibfnamefont {W.~F.}\ \bibnamefont {Vinen}},\ eds.,\ \href
  {\doibase 10.1007/3-540-45542-6} {\emph {\bibinfo {title} {{Quantized Vortex
  Dynamics and Superfluid Turbulence}}}}\ (\bibinfo  {publisher} {Springer,
  Berlin},\ \bibinfo {year} {2001})\BibitemShut {NoStop}%
\bibitem [{\citenamefont {Wlazlowski}(2020)}]{W-SLDA}%
  \BibitemOpen
  \bibfield  {author} {\bibinfo {author} {\bibfnamefont {G.}~\bibnamefont
  {Wlazlowski}},\ }\href@noop {} {\enquote {\bibinfo {title} {{Link for the
  W-SLDA package, htpp://wslda.fizyka.pw.edu.pl}},}\ } (\bibinfo {year}
  {2020})\BibitemShut {NoStop}%
\bibitem [{\citenamefont {Magierski}\ \emph {et~al.}(2009)\citenamefont
  {Magierski}, \citenamefont {Wlaz\l{}owski}, \citenamefont {Bulgac},\ and\
  \citenamefont {Drut}}]{Magierski:2009}%
  \BibitemOpen
  \bibfield  {author} {\bibinfo {author} {\bibfnamefont {P.}~\bibnamefont
  {Magierski}}, \bibinfo {author} {\bibfnamefont {G.}~\bibnamefont
  {Wlaz\l{}owski}}, \bibinfo {author} {\bibfnamefont {A.}~\bibnamefont
  {Bulgac}}, \ and\ \bibinfo {author} {\bibfnamefont {J.~E.}\ \bibnamefont
  {Drut}},\ }\bibfield  {title} {\enquote {\bibinfo {title} {Finite-temperature
  pairing gap of a unitary fermi gas by quantum monte carlo calculations},}\
  }\href {\doibase 10.1103/PhysRevLett.103.210403} {\bibfield  {journal}
  {\bibinfo  {journal} {Phys. Rev. Lett.}\ }\textbf {\bibinfo {volume} {103}},\
  \bibinfo {pages} {210403} (\bibinfo {year} {2009})}\BibitemShut {NoStop}%
\bibitem [{\citenamefont {Magierski}\ \emph {et~al.}(2011)\citenamefont
  {Magierski}, \citenamefont {Wlaz\l{}owski},\ and\ \citenamefont
  {Bulgac}}]{Magierski:2011}%
  \BibitemOpen
  \bibfield  {author} {\bibinfo {author} {\bibfnamefont {P.}~\bibnamefont
  {Magierski}}, \bibinfo {author} {\bibfnamefont {G.}~\bibnamefont
  {Wlaz\l{}owski}}, \ and\ \bibinfo {author} {\bibfnamefont {A.}~\bibnamefont
  {Bulgac}},\ }\bibfield  {title} {\enquote {\bibinfo {title} {Onset of a
  pseudogap regime in ultracold fermi gases},}\ }\href {\doibase
  10.1103/PhysRevLett.107.145304} {\bibfield  {journal} {\bibinfo  {journal}
  {Phys. Rev. Lett.}\ }\textbf {\bibinfo {volume} {107}},\ \bibinfo {pages}
  {145304} (\bibinfo {year} {2011})}\BibitemShut {NoStop}%
\bibitem [{\citenamefont {Wlaz\l{}owski}\ \emph
  {et~al.}(2013{\natexlab{a}})\citenamefont {Wlaz\l{}owski}, \citenamefont
  {Magierski}, \citenamefont {Bulgac},\ and\ \citenamefont
  {Roche}}]{Wlazlowski:2013}%
  \BibitemOpen
  \bibfield  {author} {\bibinfo {author} {\bibfnamefont {G.}~\bibnamefont
  {Wlaz\l{}owski}}, \bibinfo {author} {\bibfnamefont {P.}~\bibnamefont
  {Magierski}}, \bibinfo {author} {\bibfnamefont {A.}~\bibnamefont {Bulgac}}, \
  and\ \bibinfo {author} {\bibfnamefont {K.~J.}\ \bibnamefont {Roche}},\
  }\bibfield  {title} {\enquote {\bibinfo {title} {{Temperature evolution of
  the shear viscosity in a unitary Fermi gas}},}\ }\href {\doibase
  10.1103/PhysRevA.88.013639} {\bibfield  {journal} {\bibinfo  {journal} {Phys.
  Rev. A}\ }\textbf {\bibinfo {volume} {88}},\ \bibinfo {pages} {013639}
  (\bibinfo {year} {2013}{\natexlab{a}})}\BibitemShut {NoStop}%
\bibitem [{\citenamefont {Wlaz\l{}owski}\ \emph
  {et~al.}(2013{\natexlab{b}})\citenamefont {Wlaz\l{}owski}, \citenamefont
  {Magierski}, \citenamefont {Drut}, \citenamefont {Bulgac},\ and\
  \citenamefont {Roche}}]{Wlazlowski:2013a}%
  \BibitemOpen
  \bibfield  {author} {\bibinfo {author} {\bibfnamefont {G.}~\bibnamefont
  {Wlaz\l{}owski}}, \bibinfo {author} {\bibfnamefont {P.}~\bibnamefont
  {Magierski}}, \bibinfo {author} {\bibfnamefont {J.~E.}\ \bibnamefont {Drut}},
  \bibinfo {author} {\bibfnamefont {A.}~\bibnamefont {Bulgac}}, \ and\ \bibinfo
  {author} {\bibfnamefont {K.~J.}\ \bibnamefont {Roche}},\ }\bibfield  {title}
  {\enquote {\bibinfo {title} {{Cooper Pairing Above the Critical Temperature
  in a Unitary Fermi Gas}},}\ }\href {\doibase 10.1103/PhysRevLett.110.090401}
  {\bibfield  {journal} {\bibinfo  {journal} {Phys. Rev. Lett.}\ }\textbf
  {\bibinfo {volume} {110}},\ \bibinfo {pages} {090401} (\bibinfo {year}
  {2013}{\natexlab{b}})}\BibitemShut {NoStop}%
\bibitem [{\citenamefont {Richie-Halford}\ \emph {et~al.}(2020)\citenamefont
  {Richie-Halford}, \citenamefont {Drut},\ and\ \citenamefont
  {Bulgac}}]{Richie-Halford:2020}%
  \BibitemOpen
  \bibfield  {author} {\bibinfo {author} {\bibfnamefont {A.}~\bibnamefont
  {Richie-Halford}}, \bibinfo {author} {\bibfnamefont {J.~E.}\ \bibnamefont
  {Drut}}, \ and\ \bibinfo {author} {\bibfnamefont {A.}~\bibnamefont
  {Bulgac}},\ }\bibfield  {title} {\enquote {\bibinfo {title} {Emergence of a
  pseudogap in the bcs-bec crossover},}\ }\href {\doibase
  10.1103/PhysRevLett.125.060403} {\bibfield  {journal} {\bibinfo  {journal}
  {Phys. Rev. Lett.}\ }\textbf {\bibinfo {volume} {125}},\ \bibinfo {pages}
  {060403} (\bibinfo {year} {2020})}\BibitemShut {NoStop}%
\bibitem [{\citenamefont {Wlaz\l{}owski}\ \emph
  {et~al.}(2015{\natexlab{b}})\citenamefont {Wlaz\l{}owski}, \citenamefont
  {Quan},\ and\ \citenamefont {Bulgac}}]{Wlazlowski:2015a}%
  \BibitemOpen
  \bibfield  {author} {\bibinfo {author} {\bibfnamefont {G.}~\bibnamefont
  {Wlaz\l{}owski}}, \bibinfo {author} {\bibfnamefont {W.}~\bibnamefont {Quan}},
  \ and\ \bibinfo {author} {\bibfnamefont {A.}~\bibnamefont {Bulgac}},\
  }\bibfield  {title} {\enquote {\bibinfo {title} {{Perfect-fluid behavior of a
  dilute Fermi gas near unitary}},}\ }\href {\doibase
  10.1103/PhysRevA.92.063628} {\bibfield  {journal} {\bibinfo  {journal} {Phys.
  Rev. A}\ }\textbf {\bibinfo {volume} {92}},\ \bibinfo {pages} {063628}
  (\bibinfo {year} {2015}{\natexlab{b}})}\BibitemShut {NoStop}%
\bibitem [{\citenamefont {Bulgac}\ \emph {et~al.}(2006)\citenamefont {Bulgac},
  \citenamefont {Drut},\ and\ \citenamefont {Magierski}}]{Bulgac:2006}%
  \BibitemOpen
  \bibfield  {author} {\bibinfo {author} {\bibfnamefont {A.}~\bibnamefont
  {Bulgac}}, \bibinfo {author} {\bibfnamefont {J.~E.}\ \bibnamefont {Drut}}, \
  and\ \bibinfo {author} {\bibfnamefont {P.}~\bibnamefont {Magierski}},\
  }\bibfield  {title} {\enquote {\bibinfo {title} {{Spin $1/2$ Fermions in the
  Unitary Regime: A Superfluid of a New Type}},}\ }\href {\doibase
  10.1103/PhysRevLett.96.090404} {\bibfield  {journal} {\bibinfo  {journal}
  {Phys. Rev. Lett.}\ }\textbf {\bibinfo {volume} {96}},\ \bibinfo {pages}
  {090404} (\bibinfo {year} {2006})}\BibitemShut {NoStop}%
\bibitem [{\citenamefont {Bulgac}\ \emph {et~al.}(2008)\citenamefont {Bulgac},
  \citenamefont {Drut},\ and\ \citenamefont {Magierski}}]{Bulgac:2008a}%
  \BibitemOpen
  \bibfield  {author} {\bibinfo {author} {\bibfnamefont {A.}~\bibnamefont
  {Bulgac}}, \bibinfo {author} {\bibfnamefont {J.~E.}\ \bibnamefont {Drut}}, \
  and\ \bibinfo {author} {\bibfnamefont {P.}~\bibnamefont {Magierski}},\
  }\bibfield  {title} {\enquote {\bibinfo {title} {{Quantum Monte Carlo
  simulations of the BCS-BEC crossover at finite temperature}},}\ }\href
  {\doibase 10.1103/PhysRevA.78.023625} {\bibfield  {journal} {\bibinfo
  {journal} {Phys. Rev. A}\ }\textbf {\bibinfo {volume} {78}},\ \bibinfo
  {pages} {023625} (\bibinfo {year} {2008})}\BibitemShut {NoStop}%
\bibitem [{\citenamefont {Drut}\ \emph {et~al.}(2012)\citenamefont {Drut},
  \citenamefont {L\"ahde}, \citenamefont {Wlaz\l{}owski},\ and\ \citenamefont
  {Magierski}}]{Drut:2012}%
  \BibitemOpen
  \bibfield  {author} {\bibinfo {author} {\bibfnamefont {J.~E.}\ \bibnamefont
  {Drut}}, \bibinfo {author} {\bibfnamefont {T.~A.}\ \bibnamefont {L\"ahde}},
  \bibinfo {author} {\bibfnamefont {G.}~\bibnamefont {Wlaz\l{}owski}}, \ and\
  \bibinfo {author} {\bibfnamefont {P.}~\bibnamefont {Magierski}},\ }\bibfield
  {title} {\enquote {\bibinfo {title} {{Equation of state of the unitary Fermi
  gas: An update on lattice calculations}},}\ }\href {\doibase
  10.1103/PhysRevA.85.051601} {\bibfield  {journal} {\bibinfo  {journal} {Phys.
  Rev. A}\ }\textbf {\bibinfo {volume} {85}},\ \bibinfo {pages} {051601}
  (\bibinfo {year} {2012})}\BibitemShut {NoStop}%
\bibitem [{\citenamefont {Hahn}\ and\ \citenamefont
  {Strassmann}(1939)}]{Hahn:1939}%
  \BibitemOpen
  \bibfield  {author} {\bibinfo {author} {\bibfnamefont {O.}~\bibnamefont
  {Hahn}}\ and\ \bibinfo {author} {\bibfnamefont {F.}~\bibnamefont
  {Strassmann}},\ }\bibfield  {title} {\enquote {\bibinfo {title} {{\"Uber den
  Nachweis und das Verhalten der bei der Bestrahlung des Urans mittels
  Neutronen entstehenden Erdalkalimetalle}},}\ }\href {\doibase
  10.1007/BF01488241} {\bibfield  {journal} {\bibinfo  {journal}
  {Naturwissenschaften}\ }\textbf {\bibinfo {volume} {27}},\ \bibinfo {pages}
  {11} (\bibinfo {year} {1939})}\BibitemShut {NoStop}%
\bibitem [{\citenamefont {Bohr}(1936)}]{Bohr:1936}%
  \BibitemOpen
  \bibfield  {author} {\bibinfo {author} {\bibfnamefont {N.}~\bibnamefont
  {Bohr}},\ }\bibfield  {title} {\enquote {\bibinfo {title} {{Neutron Capture
  and Nuclear Constitution}},}\ }\href {\doibase doi.org/10.1038/137344a0}
  {\bibfield  {journal} {\bibinfo  {journal} {Nature}\ }\textbf {\bibinfo
  {volume} {137}},\ \bibinfo {pages} {344} (\bibinfo {year}
  {1936})}\BibitemShut {NoStop}%
\bibitem [{\citenamefont {{News and Views (editorial)}}(1936)}]{Bohr:1936a}%
  \BibitemOpen
  \bibfield  {author} {\bibinfo {author} {\bibnamefont {{News and Views
  (editorial)}}},\ }\bibfield  {title} {\enquote {\bibinfo {title} {{Neutron
  Capture and Nuclear Constitution}},}\ }\href {\doibase
  https://doi.org/10.1038/137351b0} {\bibfield  {journal} {\bibinfo  {journal}
  {Nature}\ }\textbf {\bibinfo {volume} {137}},\ \bibinfo {pages} {351}
  (\bibinfo {year} {1936})}\BibitemShut {NoStop}%
\bibitem [{\citenamefont {Meitner}\ and\ \citenamefont
  {Frisch}(1939)}]{Meitner:1939}%
  \BibitemOpen
  \bibfield  {author} {\bibinfo {author} {\bibfnamefont {L.}~\bibnamefont
  {Meitner}}\ and\ \bibinfo {author} {\bibfnamefont {O.~R.}\ \bibnamefont
  {Frisch}},\ }\bibfield  {title} {\enquote {\bibinfo {title} {{Disintegration
  of Uranium by Neutrons: a New Type of Nuclear Reaction}},}\ }\href
  {http://dx.doi.org/10.1038/143239a0} {\bibfield  {journal} {\bibinfo
  {journal} {Nature}\ }\textbf {\bibinfo {volume} {143}},\ \bibinfo {pages}
  {239} (\bibinfo {year} {1939})}\BibitemShut {NoStop}%
\bibitem [{\citenamefont {Fermi}(1949)}]{Fermi:1949}%
  \BibitemOpen
  \bibfield  {author} {\bibinfo {author} {\bibfnamefont {E.}~\bibnamefont
  {Fermi}},\ }\bibfield  {title} {\enquote {\bibinfo {title} {On the origin of
  the cosmic radiation},}\ }\href {\doibase 10.1103/PhysRev.75.1169} {\bibfield
   {journal} {\bibinfo  {journal} {Phys. Rev.}\ }\textbf {\bibinfo {volume}
  {75}},\ \bibinfo {pages} {1169} (\bibinfo {year} {1949})}\BibitemShut
  {NoStop}%
\bibitem [{\citenamefont {Balian}\ and\ \citenamefont
  {{V\'en\'eroni}}(1984)}]{Balian:1984}%
  \BibitemOpen
  \bibfield  {author} {\bibinfo {author} {\bibfnamefont {R.}~\bibnamefont
  {Balian}}\ and\ \bibinfo {author} {\bibfnamefont {M.}~\bibnamefont
  {{V\'en\'eroni}}},\ }\bibfield  {title} {\enquote {\bibinfo {title}
  {Fluctuations in a time-dependent mean-field approach},}\ }\href {\doibase
  10.1016/0370-2693(84)92008-2} {\bibfield  {journal} {\bibinfo  {journal}
  {Phys. Lett. B}\ }\textbf {\bibinfo {volume} {136}},\ \bibinfo {pages} {301}
  (\bibinfo {year} {1984})}\BibitemShut {NoStop}%
\bibitem [{\citenamefont {Balian}\ \emph {et~al.}(1984)\citenamefont {Balian},
  \citenamefont {Bonche}, \citenamefont {Flocard},\ and\ \citenamefont
  {V\'en\'eroni}}]{Balian:1984a}%
  \BibitemOpen
  \bibfield  {author} {\bibinfo {author} {\bibfnamefont {R.}~\bibnamefont
  {Balian}}, \bibinfo {author} {\bibfnamefont {P.}~\bibnamefont {Bonche}},
  \bibinfo {author} {\bibfnamefont {H.}~\bibnamefont {Flocard}}, \ and\
  \bibinfo {author} {\bibfnamefont {M.}~\bibnamefont {V\'en\'eroni}},\
  }\bibfield  {title} {\enquote {\bibinfo {title} {Mass dispersions in a
  time-dependent mean-field approach},}\ }\href {\doibase
  10.1016/0375-9474(84)90243-4} {\bibfield  {journal} {\bibinfo  {journal}
  {Nucl. Phys. A}\ }\textbf {\bibinfo {volume} {428}},\ \bibinfo {pages} {79}
  (\bibinfo {year} {1984})}\BibitemShut {NoStop}%
\bibitem [{\citenamefont {Balian}\ and\ \citenamefont
  {V\'en\'eroni}(1985)}]{Balian:1985}%
  \BibitemOpen
  \bibfield  {author} {\bibinfo {author} {\bibfnamefont {R.}~\bibnamefont
  {Balian}}\ and\ \bibinfo {author} {\bibfnamefont {M.}~\bibnamefont
  {V\'en\'eroni}},\ }\bibfield  {title} {\enquote {\bibinfo {title}
  {{Time-Dependent Variational Principle for the Expectation Value of an
  Observable: Mean-Field Applications}},}\ }\href {\doibase
  10.1016/0003-4916(85)90020-X} {\bibfield  {journal} {\bibinfo  {journal}
  {Ann. Phys.}\ }\textbf {\bibinfo {volume} {164}},\ \bibinfo {pages} {334}
  (\bibinfo {year} {1985})}\BibitemShut {NoStop}%
\bibitem [{\citenamefont {Simenel}(2011)}]{Simenel:2011}%
  \BibitemOpen
  \bibfield  {author} {\bibinfo {author} {\bibfnamefont {C.}~\bibnamefont
  {Simenel}},\ }\bibfield  {title} {\enquote {\bibinfo {title}
  {{Particle-Number Fluctuations and Correlations in Transfer Reactions
  Obtained Using the Balian-V\'en\'eroni Variational Principle}},}\ }\href
  {\doibase 10.1103/PhysRevLett.106.112502} {\bibfield  {journal} {\bibinfo
  {journal} {Phys. Rev. Lett.}\ }\textbf {\bibinfo {volume} {106}},\ \bibinfo
  {pages} {112502} (\bibinfo {year} {2011})}\BibitemShut {NoStop}%
\bibitem [{\citenamefont {Simenel}(2012)}]{Simenel:2012}%
  \BibitemOpen
  \bibfield  {author} {\bibinfo {author} {\bibfnamefont {C.}~\bibnamefont
  {Simenel}},\ }\bibfield  {title} {\enquote {\bibinfo {title} {Nuclear quantum
  many-body dynamics},}\ }\href {\doibase 10.1140/epja/i2012-12152-0}
  {\bibfield  {journal} {\bibinfo  {journal} {Eur. Phys. Jour. A}\ }\textbf
  {\bibinfo {volume} {48}},\ \bibinfo {pages} {152} (\bibinfo {year}
  {2012})}\BibitemShut {NoStop}%
\bibitem [{\citenamefont {Scamps}\ \emph {et~al.}(2015)\citenamefont {Scamps},
  \citenamefont {Simenel},\ and\ \citenamefont {Lacroix}}]{Scamps:2015}%
  \BibitemOpen
  \bibfield  {author} {\bibinfo {author} {\bibfnamefont {G.}~\bibnamefont
  {Scamps}}, \bibinfo {author} {\bibfnamefont {C.}~\bibnamefont {Simenel}}, \
  and\ \bibinfo {author} {\bibfnamefont {D.}~\bibnamefont {Lacroix}},\
  }\bibfield  {title} {\enquote {\bibinfo {title} {Superfluid dynamics of
  $^{258}\mathrm{Fm}$ fission},}\ }\href {\doibase 10.1103/PhysRevC.92.011602}
  {\bibfield  {journal} {\bibinfo  {journal} {Phys. Rev. C}\ }\textbf {\bibinfo
  {volume} {92}},\ \bibinfo {pages} {011602} (\bibinfo {year}
  {2015})}\BibitemShut {NoStop}%
\bibitem [{\citenamefont {Williams}\ \emph {et~al.}(2018)\citenamefont
  {Williams}, \citenamefont {Sekizawa}, \citenamefont {Hinde}, \citenamefont
  {Simenel}, \citenamefont {Dasgupta}, \citenamefont {Carter}, \citenamefont
  {Cook}, \citenamefont {Jeung}, \citenamefont {McNeil}, \citenamefont
  {Palshetkar}, \citenamefont {Rafferty}, \citenamefont {Ramachandran},\ and\
  \citenamefont {Wakhle}}]{Williams:2018}%
  \BibitemOpen
  \bibfield  {author} {\bibinfo {author} {\bibfnamefont {E.}~\bibnamefont
  {Williams}}, \bibinfo {author} {\bibfnamefont {K.}~\bibnamefont {Sekizawa}},
  \bibinfo {author} {\bibfnamefont {D.~J.}\ \bibnamefont {Hinde}}, \bibinfo
  {author} {\bibfnamefont {C.}~\bibnamefont {Simenel}}, \bibinfo {author}
  {\bibfnamefont {M.}~\bibnamefont {Dasgupta}}, \bibinfo {author}
  {\bibfnamefont {I.~P.}\ \bibnamefont {Carter}}, \bibinfo {author}
  {\bibfnamefont {K.~J.}\ \bibnamefont {Cook}}, \bibinfo {author}
  {\bibfnamefont {D.~Y.}\ \bibnamefont {Jeung}}, \bibinfo {author}
  {\bibfnamefont {S.~D.}\ \bibnamefont {McNeil}}, \bibinfo {author}
  {\bibfnamefont {C.~S.}\ \bibnamefont {Palshetkar}}, \bibinfo {author}
  {\bibfnamefont {D.~C.}\ \bibnamefont {Rafferty}}, \bibinfo {author}
  {\bibfnamefont {K.}~\bibnamefont {Ramachandran}}, \ and\ \bibinfo {author}
  {\bibfnamefont {A.}~\bibnamefont {Wakhle}},\ }\bibfield  {title} {\enquote
  {\bibinfo {title} {Exploring zeptosecond quantum equilibration dynamics: From
  deep-inelastic to fusion-fission outcomes in
  $^{58}\mathrm{Ni}+^{60}\mathrm{Ni}$ reactions},}\ }\href {\doibase
  10.1103/PhysRevLett.120.022501} {\bibfield  {journal} {\bibinfo  {journal}
  {Phys. Rev. Lett.}\ }\textbf {\bibinfo {volume} {120}},\ \bibinfo {pages}
  {022501} (\bibinfo {year} {2018})}\BibitemShut {NoStop}%
\bibitem [{\citenamefont {Volya}\ and\ \citenamefont
  {Zelevinsky}(2020)}]{Volya:2020}%
  \BibitemOpen
  \bibfield  {author} {\bibinfo {author} {\bibfnamefont {A.}~\bibnamefont
  {Volya}}\ and\ \bibinfo {author} {\bibfnamefont {V.}~\bibnamefont
  {Zelevinsky}},\ }\bibfield  {title} {\enquote {\bibinfo {title}
  {Time-dependent relaxation of observables in complex quantum systems},}\
  }\href {\doibase 10.1088/2632-072x/ab79bc} {\bibfield  {journal} {\bibinfo
  {journal} {Journal of Physics: Complexity}\ }\textbf {\bibinfo {volume}
  {1}},\ \bibinfo {pages} {025007} (\bibinfo {year} {2020})}\BibitemShut
  {NoStop}%
\end{thebibliography}%

\end{document}